\documentclass[12pt]{article}
\usepackage{newtxtext,newtxmath}

\usepackage{graphicx}
\usepackage[letterpaper,margin=1in]{geometry}
\renewenvironment{abstract}
	{\quotation}
	{\endquotation}
\date{}

\makeatletter
\renewcommand{\fnum@figure}{\textbf{Figure \thefigure}}
\renewcommand{\fnum@table}{\textbf{Table \thetable}}
\makeatother
\usepackage{scicite}
\usepackage{url}
\usepackage{orcidlink}

\def\scititle{
Ultrasonic metamaterial at MHz frequencies using microstructured glass
}
\title{\bfseries \boldmath \scititle}
\author{
	Oscar~Demeulenaere$^{1,2\ast\orcidlink{0000-0003-4492-5024}}$,
	Nikita~Ustimenko$^{3\orcidlink{0000-0002-5137-493X}}$,
    Athanasios~G.~Athanassiadis$^{1,2\orcidlink{0000-0002-2015-829X}}$,\and
    Lovish~Gulati$^{1,2}$,
    Carsten~Rockstuhl$^{3,4\orcidlink{0000-0002-5868-0526}}$,
    Peer~Fischer$^{1,2,5,6\orcidlink{0000-0002-8600-5958}}$,\and
    \small$^{1}$
    Max Planck Institute for Medical Research, Jahnstr. 29, 69120 Heidelberg, Germany \and
    \small$^{2}$Institute for Molecular Systems Engineering and Advanced Materials,
    \and \small Heidelberg University, Im Neuenheimer Feld 225, 69120 Heidelberg, Germany \and
    \small$^{3}$Institute of Theoretical Solid State Physics,
    \and \small Karlsruhe Institute of Technology, Kaiserstr. 12, 76131 Karlsruhe, Germany.\and
    \small$^{4}$Institute of Nanotechnology,
    \and \small Karlsruhe Institute of Technology, Kaiserstr. 12, 76131 Karlsruhe, Germany. \and
    \small$^{5}$Center for Nanomedicine, Institute for Basic Science (IBS), Seoul 03722, Republic of Korea \and
    \small$^{6}$Department of Nano Biomedical Engineering (NanoBME),
    \and \small Advanced Science Institute, Yonsei University, Seoul 03722, Republic of Korea \and
	\small$^\ast$Corresponding author. Email: oscar.demeulenaere@mr.mpg.de \and
}

\begin{document} 
\maketitle
\begin{abstract} \bfseries \boldmath
Acoustic metamaterials enhance traditional material properties, providing opportunities to shape sound fields in applications ranging from biomedical imaging and clinical therapy to non-destructive testing. However, at the MHz frequency range, only a few metamaterial architectures exist. They are often highly attenuating or difficult to manufacture and generally provide limited control. Here, we introduce a MHz-frequency ultrasonic metamaterial based on laser-engraved glass. By structuring meta-voxels with different engraving patterns, we define an anisotropic metamaterial exhibiting local variations in the sound speed of up to 20\% compared to unstructured glass, and losses $100\times$ lower than in comparable 3D-printed metamaterials. We use this metamaterial to define a library of standard elements that can be modularly combined to shape complex-patterned ultrasonic fields. Our experiments are supported by a theoretical model, which provides additional insights into the microstructural origin of the metamaterial behavior and opens the door to designing tailored ultrasound fields and responses.
\end{abstract}

\noindent
\subsection*{Introduction}
The ability to precisely shape ultrasonic wavefronts is critical for diverse biomedical and engineering applications. In medicine, MHz-frequency ultrasound offers deep tissue penetration with millimeter-scale focusing, making it indispensable for noninvasive imaging and therapeutic interventions. Wavefront shaping in these contexts enables the correction of aberrations introduced by heterogeneous media and precisely localizes ultrasonic energy for imaging or therapeutic techniques such as transcranial neuromodulation~\cite{attali_deep_2025} and thermal ablation~\cite{cengiz_roadmap_2025}. Emerging applications in acoustofluidics leverage patterned ultrasound fields for particle sorting and microassembly~\cite{xu_micro-acoustic_2024}, tissue engineering~\cite{shi_acoustic_2025}, and even 3D printing~\cite{derayatifar_holographic_2024}. Industrial testing similarly benefits from the high precision and long propagation distances of MHz-frequency ultrasound~\cite{zhao_nondestructive_2022}. In such applications, deliberate scattering masks and dispersion engineering can provide additional information about complex structures being imaged~\cite{kruizinga_compressive_2017, lee_anisotropic_2025}. Finally, approaches to physical computation can benefit from techniques to fabricate intricate three-dimensional (3D) structures for acoustic wavefront control~\cite{hughes_wave_2019}.

Typically, ultrasonic wavefronts are either actively shaped using phased arrays or passively shaped using phase-modulating elements, such as lenses or holograms~\cite{kruizinga_compressive_2017}. Ultrasonic phased arrays are the gold standard for imaging applications, but do not scale well to the large element counts needed to project complex fields, due to the increasing hardware complexity and cost of the required driving electronics. In many cases, it is more effective to control ultrasonic wavefronts using passive phase-modulating elements paired with single-element transducers. A common technique relies on phase plates fabricated from polymers whose speed of sound differs from that in the surrounding (typically water-like) medium. By appropriately designing the phase plate thickness, a phase shift can be introduced to the impinging wavefront, giving rise to a complex pressure field pattern in the desired focal region~\cite{melde_holograms_2016}. However, polymeric phase elements typically exhibit high absorptive losses that introduce limitations for use with high-power sources~\cite{burstow_acoustic_2025} or for use in complex field shaping using 3D volumetric holograms~\cite{brown_binary_2023}.

An alternative approach for wavefront engineering that offers even more degrees of freedom is to use acoustic metamaterials. Metamaterials are microstructured materials comprised of subwavelength unit cells, whose collective influence on an incident wave gives rise to diverse properties, including many that cannot be realized in traditional homogeneous materials. In addition to providing tunable speeds of sound, metamaterials allow wave propagation to be designed in more complex ways, including anisotropic (e.g., birefringent), dispersive (frequency-dependent), and topologically-guided behaviors~\cite{naify_what_2024, cummer_controlling_2016}. Despite such promise, however, the availability of metamaterial architectures that function at MHz frequencies is limited by challenges in fabrication and scaling. Conventional acoustic metamaterial architectures, such as elastomer composites~\cite{qu_underwater_2022, guillermic_pdms-based_2019}, labyrinthine structures, micro-pillars, or Helmholtz resonators~\cite{assouar_acoustic_2018}, form a powerful toolbox for engineering low-frequency devices, but have not been effectively scaled for use at MHz frequencies. Experimental realizations in the MHz regime include silicone-based gradient-index metasurfaces~\cite{jin_flat_2019} for broadband focusing and beam steering, and lithographically-manufactured planarized micropillar arrays that precisely modulate the phase shift across a thin structure for acoustic holography~\cite{xu_sound-speed_2023}. While most metamaterials explored have been two-dimensional (2D) metastructures~\cite{xu_sound-speed_2023,Dorn2026Feb,Olsson2008,jin_flat_2019,melde_holograms_2016,attali_deep_2025,Melnikov2022Jul}, a growing
number of 3D metamaterials have recently emerged, leveraging advances
in 3D printing~\cite{brown_binary_2023,IglesiasMartinez2021Feb}. 3D metamaterial architectures expand the possibilities for field control across scales. However, their reliance on UV-cured polymers limits the mechanical and thermal stability of such materials, and also introduces very high attenuation coefficients that constrain the practical device size. A detailed comparison of existing 3D and 2D metamaterials for MHz ultrasonic waves is given in Tables~\ref{tab:table_literature_3d} and~\ref{tab:table_literature_2d}, respectively. 

However, such current approaches face several limitations. While microfabrication techniques are suitable for small devices, the fabrication of accurate structures to modulate large-area sources (e.g., $>5$ cm), such as those relevant in many higher-power applications, remains limited by manufacturing costs and alignment errors. Another constraint arises when the material in use exhibits high acoustic losses at MHz frequencies. 3D-printed photopolymers such as VeroClear (Stratasys Inc) that are typically used for 3D-printed phase plates have an attenuation of 370 dB\,MHz$^{-1}$\,m$^{-1}$ at 2 MHz~\cite{bakaric_measurement_2021}, which is orders of magnitude higher than that of other typical ultrasonic propagation materials (Table~\ref{tab:tab1}). Other metamaterials based on silicone similarly exhibit high attenuation at MHz frequencies~\cite{itsumi_low_2009}. Beyond just limiting power delivery to the target, high attenuation introduces heating that can lead to detrimental effects in high-power applications, such as variations in speed of sound and geometric warping, which are known to reduce the performance of phase plates~\cite{andres_holographic_2023}.

To overcome such limitations, we introduce here a low-loss and scalable metamaterial architecture based on microstructured glass, capable of shaping ultrasound wavefronts at MHz frequencies. By using glass as the underlying substrate, our metamaterial architecture benefits from the extremely low ultrasonic losses~\cite{ono_comprehensive_2020}, high structural stability, and high thermal stability of glass. We pattern a defined microstructure into BK7 glass using subsurface laser engraving (see Fig.~\ref{fig:fig1}A) -- a 3D laser-processing technique that creates individual, 100-\textmu m scale microcracks at precisely defined locations within the glass. To produce the cracks, a high-peak-power pulsed laser is focused into the bulk of the glass, where optical breakdown -- a nonlinear optical process -- locally fractures the material without damaging the substrate away from the optical focus~\cite{Kruger1998May}. Depending on laser pulse length and energy, different crack morphologies can be produced: ultrashort (femto- to picosecond), lower-energy pulses produce small and smoother cracks, while longer (nanosecond), higher-energy pulses produce larger and rougher cracks~\cite{Glezer1997Aug, Glezer1997May}. To ensure sufficiently strong scattering of MHz acoustic waves to support metamaterial behavior, we chose to use a standard, commercially accessible micromachining technique based on a nanosecond laser (pulse duration: 9~ns), which produces 100-micron-scale cracks. Such an approach has been previously explored to manufacture artificial flaws for calibration of non-destructive testing probes and has been shown to exhibit scattering and shear-wave birefringence in the MHz frequency range~\cite{szelazek_application_2014}. Here, we substantially build upon these earlier results to pattern arbitrarily shaped metamaterial regions within large (centimeter-scale) glass volumes to manipulate ultrasonic waves in 3D.

We characterize the effective properties of a laser-engraved voxel consisting of thousands of microcracks whose sizes and spacings are much smaller than the wavelength. We measured the effective speed of sound and density of one voxel using a refracto-vibrometry technique. Notably, we observed that the effective parameters are strongly anisotropic, which we relate to the microcrack geometry. The response is also highly frequency-dependent, pointing to resonant interactions within the metamaterial. We introduce and experimentally validate a numerical transition-matrix (\textit{T}-matrix) model~\cite{Waterman1969Jun,Ustimenko2025Apr,Ustimenko2025Oct} describing the metamaterial, which allows the scattering properties of an arbitrary pattern of microcracks to be accurately predicted based on the scattering characteristics of a single microcrack. This \textit{T}-matrix approach also confirms the metamaterial nature of the engraved material.

Building upon the characterization of the metamaterial unit cells, we designed distinct elements to control the ultrasonic wavefront inside the glass: a lens, a grating, a waveguide, and several acoustic holograms. All of these elements exhibit very low attenuation in the MHz frequency range. Finally, we highlight the modularity of this platform by combining individual metamaterial elements to create more complex acoustic wavefront-shaping systems, reminiscent of complex optical assemblies based on fundamental elements.

\subsection*{Results}
\subsubsection*{Micro-engraved glass constitutes a tunable MHz-frequency metamaterial}

\begin{figure}[h!]
	\centering
	\includegraphics[width=0.9\textwidth]{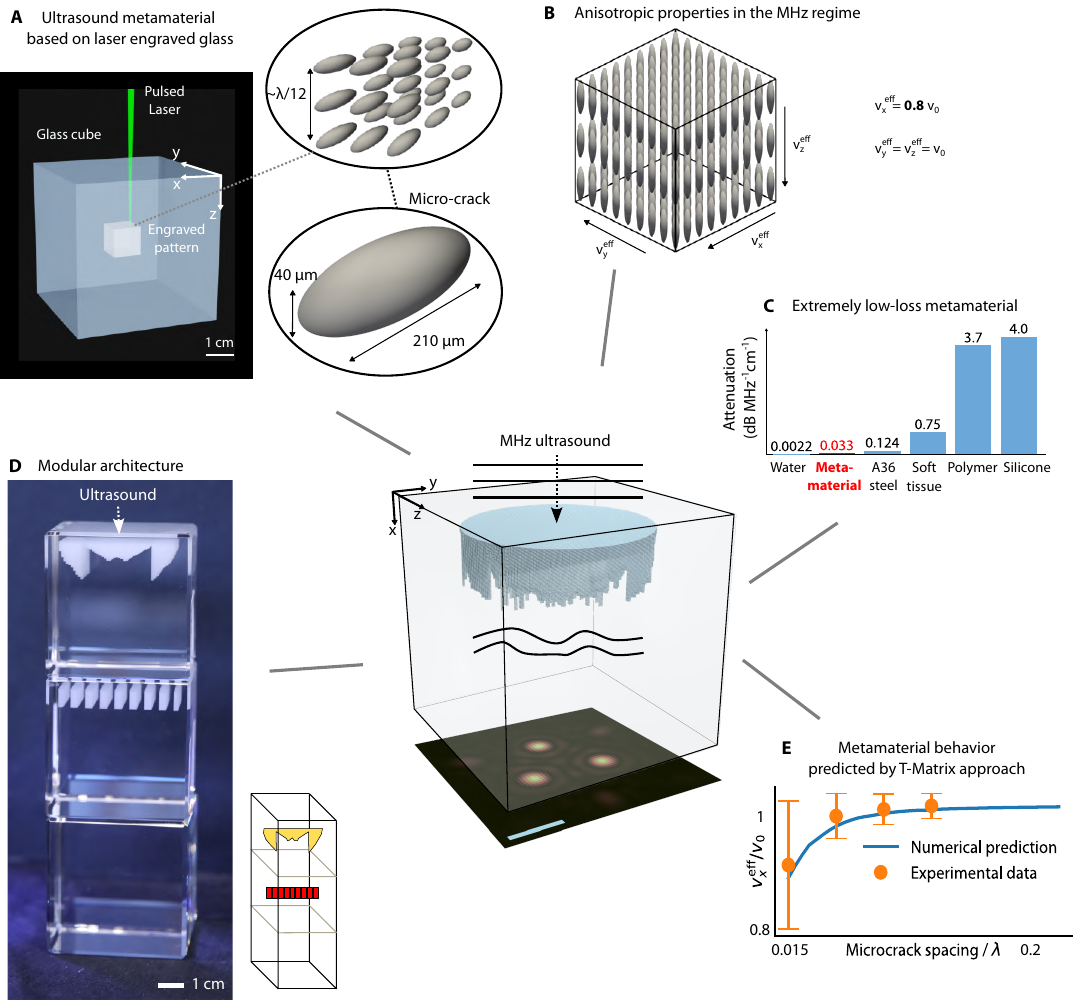}
	\caption{\textbf{Overview of a metamaterial architecture for MHz-frequency ultrasound based on microstructured glass.}
    We introduce a glass metamaterial, which can be spatially patterned in 3D for complex wavefront engineering. For example, a three-foci acoustic hologram can be generated from an MHz-ultrasonic wavefront (main panel) by patterning subwavelength laser-induced microcracks in regular grids (panel \textbf{A}). (\textbf{B}) The effective speed of sound and density of the glass are then modified along only one propagation direction ($x$-axis). (\textbf{C}) In comparison with other common metamaterials, this glass metamaterial (in red) exhibits extremely low attenuation at MHz frequencies. (\textbf{D}) The architecture is modular, and the different blocks can be combined to produce wavefront-shaping assemblies. (\textbf{E}) A \textit{T}-matrix framework can also provide numerical predictions of the metamaterial behavior. All scale bars indicate 1 cm.}
	\label{fig:fig1}
\end{figure}

To create our metamaterials, 3D patterns of microcracks were machined inside 5~cm glass cubes using a commercially available subsurface laser engraving service (M+B Lasertechnik GmbH). In this process, a 532~nm pulsed laser is precisely focused onto user-defined positions within the glass using computer-controlled adjustable lenses. It produces micro-fractures without damaging the bulk medium (Fig.~\ref{fig:fig1}). For our most commonly used machining parameters, the microfractures have typical diameters perpendicular to the machining laser axis ($x$- and $y$-axes) of $32\pm1$ $\mu$m along the $x$-axis, $19\pm1$ $\mu$m along the $y$-axis, and lengths of $231\pm2$ $\mu$m along the laser axis ($z$-axis). The geometry of the microcracks was measured from light microscopy images of three samples with the same engraving parameters. The coefficients of variation among the samples were 3.6\%, 6.0\%, and 0.9\% for the $x$-, $y$-, and $z$-axes, respectively, indicating good fabrication reproducibility.

By patterning the cracks, a tunable phase shift could be introduced to an incident compressional wave in the glass due to a change in the effective sound speed of the micropatterned region. The unmodified glass blocks have a  density of $\rho_{0}=2525\pm8$ kg\,m$^{-3}$ and a longitudinal speed of sound of $v_{0}=5816\pm4$~m\,s$^{-1}$ (see Materials and Methods for details on their measurement). The values are in good agreement with the literature for BK7 glass, which is a typical glass material used in subsurface laser engraving~\cite{ono_comprehensive_2020}.
These baseline acoustic properties were then compared with the effective sound speed and density of rectangular meta-voxels within an otherwise homogeneous glass block. Meta-voxels consisted of $45 \times 45 \times 115$ microcracks arranged in a 3D rectangular grid spanning $4 \times 4 \times 31$ mm$^3$.  

\begin{figure}
	\centering
	\includegraphics[width=1\textwidth]{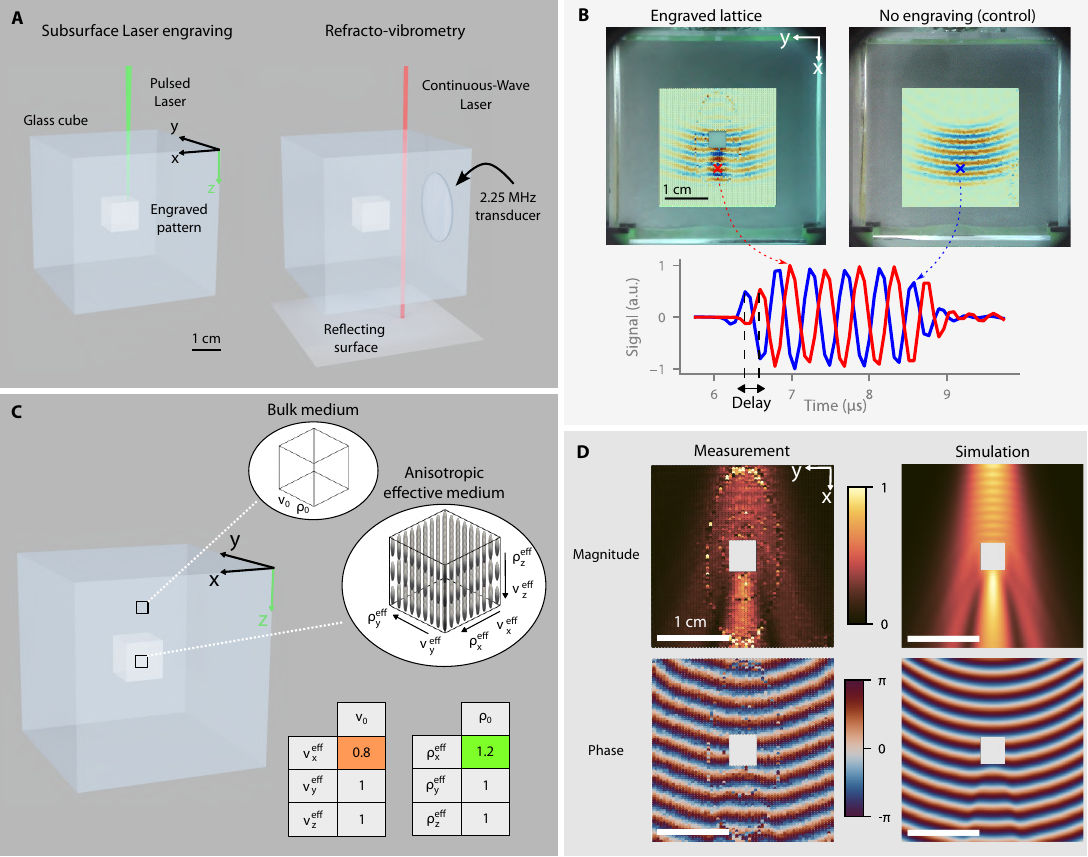}
	\caption{\textbf{Effective medium characterization.}
    (\textbf{A}) Schematics of the subsurface laser engraving process (left) and of the refracto-vibrometry measurements (right). The $z$-axis is chosen as the engraving direction of the laser and depicted in green. (\textbf{B}) Refracto-vibrometry measurements show that a five-cycle burst at 2.25 MHz is delayed when propagating through an engraved voxel compared to a control block (without any engraving). (\textbf{C}) From the time-of-flight calculations, the effective speed of sound of the engraved area was determined for every propagation direction. From the effective speed of sound, the effective density was retrieved. (\textbf{D}) \textit{T}-matrix simulations show a good agreement with the refracto-vibrometry measurements. All scale bars indicate 1 cm.}
	\label{fig:fig2}
\end{figure}

To measure the effective medium properties, pulsed compressional waves with a center frequency of $f_{0} = 2.25$~MHz were incident on the sample, and the wave propagation within the glass block was observed using refracto-vibrometry (Fig.~\ref{fig:fig2}A). By combining the known length of the metavoxel and the measured transit time of the pulse through the metavoxel (Fig.~\ref{fig:fig2}B), we were able to measure the effective speed of compressional waves in the metavoxel. Then, using this effective speed of sound and together with the reflected-wave amplitude from the metavoxel, we could determine its effective density. These measurements were repeated for incident waves along all three principal axes of the metavoxel ($x$, $y$, and $z$), where the long axis of the voxel was always parallel to the vibrometry axis (see Materials and Methods). 

These measurements revealed that the patterned microcracks reduce the effective speed of sound in the glass and increase the effective density by up to 20\% (Fig.~\ref{fig:fig2}C and Table~\ref{tab:tab2}). This effect is, however, anisotropic and only appears along the $x$-axis. From six different samples with identical engraving parameters, the measured  effective speed of sound and density showed a relative deviation of less than 4\%. The engraving process is thus reproducible across multiple samples.

While a difference in effective medium properties in the $z$-direction is expected because of the extended crack length along the laser propagation direction, the difference between the $x$- and $y$-directions is surprising because of the rotational symmetry of the laser beam. To understand the origin of this birefringent behavior, we studied the microcrack geometry using light microscopy. We observed that the microcracks do not have the same geometry along the $x$- and $y$-axes (Fig.~\ref{fig:SI_Fig_1}). This asymmetry is nonetheless consistent across cracks and samples, suggesting a reproducible instability in the laser–matter interaction at the focal point. The microcracks are indeed elongated along the $z$-axis due to plasma growth toward the laser source and anisotropic in the $xy$-plane, possibly because the laser's linear polarization creates an asymmetric stress field~\cite{Rastogi2024Aug}.

Using the crack geometry as a constituent scatterer, we built a numerical model to explore the metamaterial behavior more generally using the \textit{T}-matrix approach. In short, the linear scattering behavior of a single microcrack was described analytically in the frequency domain, assuming a subwavelength ellipsoidal scatterer with anisotropic material parameters located in the unbounded, homogeneous medium, which is the bulk glass medium (see Materials and Methods). In this regard, the acoustic response of the microcrack is completely determined by its \textit{T}-matrix, allowing us to predict wave propagation behavior through arbitrary arrangements of cracks while accounting for multiple scattering and collective interactions between cracks.

The scattered field predicted by this method for a 3D grid of microcracks shows excellent agreement with the refracto-vibrometry data (Fig.~\ref{fig:fig2}D). As these 2D vibrometry measurements are projections of the 3D acoustic field inside the glass, we applied a quasi-averaging procedure to the numerical results to reproduce the expected measurement data. In addition to the pressure-field pattern, the effective speed of sound and density from the simulation exhibit excellent quantitative agreement with the experimental results (Fig.~\ref{fig:SI_Fig_2}). The \textit{T}-matrix simulation is also able to predict the effective sound speed as a function of crack spacing (Figs.~\ref{fig:fig1}D and~\ref{fig:SI_Fig_3}). Moreover, the \textit{T}-matrix predictions accurately capture the anisotropic properties of the metavoxel using only a monopole--dipole approximation, revealing that the metamaterial behavior arises primarily from the strongly anisotropic dipole response of a single microcrack.

The combined experimental and numerical results for the metavoxel speed of sound demonstrate the metamaterial nature of the microstructured glass: the strong scattering behavior of individual deeply subwavelength microcracks gives rise to an emergent effective medium at wavelength-scales and above. As commonly observed in metamaterials, the effective-medium properties of the microstructured glass metavoxels were also revealed to be highly frequency-dependent. A difference between the speed of sound in the bulk material and that of the effective medium is observed only within a frequency window centered at approximately 3~MHz. Experimentally, a full width at half maximum (FWHM) of 3.1~MHz was measured, which corresponds to a fractional bandwidth of 87\%. Below 0.5~MHz and above 4~MHz, no measurable difference was observed between the two media (Fig.~\ref{fig:SI_Fig_4}).

Finally, we measured the attenuation of compressional waves in the metavoxel and found that the attenuation coefficient of the microstructured glass metamaterial is exceptionally low. Whereas the bulk glass exhibits an attenuation of $\alpha = 3.3 \pm 0.4$~dB\,m$^{-1}$\,MHz$^{-1}$, in good agreement with previous reports~\cite{ono_comprehensive_2020}, we find the microstructured glass to exhibit isotropic attenuation only 10\% higher than the bulk glass (Table~\ref{tab:tab2} and Fig.~\ref{fig:SI_Fig_5}). For comparison, polymers and other materials commonly used for ultrasonic wavefront shaping exhibit attenuation coefficients that are $100\times$ higher in the MHz frequency range (Table~\ref{tab:tab1}). 

\subsubsection*{Microstructured glass metamaterial can be patterned for modular, arbitrary wavefront shaping}

Because of its ease of manufacturing, control over speed of sound, and low losses, we use our metamaterial to create a library of wavefront-shaping elements made of glass. All elements were laser engraved into $5 \times 5 \times 5$ cm$^3$ glass blocks (Fig.~\ref{fig:fig3}). To illustrate the variety of elements that can be achieved, we manufactured a focusing lens, a grating, a waveguide, and multi-focus holograms. Using both refracto-vibrometry through the glass and surface vibrometry at the face opposite the transducer, we observed excellent control over the acoustic wave propagating inside the glass at 2.25~MHz.

Among the benefits of using glass for wavefront shaping are its high structural and thermal robustness (Table~\ref{tab:table_s1}) and its corresponding ability to withstand high acoustic powers. Tests of our elements in acoustic fields of increasing intensity revealed no structural damage or wavefront aberration with increasing intensity, up to a maximum intensity of $20~\mathrm{W\,cm}^{-2}$ that our transducer could transmit.

Another major benefit of glass-block meta-elements is their ability to couple sound effectively through their flat faces. Beyond general coupling, this opens the door to modularity by stacking and rearranging elements face-to-face (Fig.~\ref{fig:fig4}). To demonstrate these strengths, we sequentially composed different wavefront transformations to realize more complex manipulations with acoustic wavefronts. For instance, coupling a Fresnel lens block to a grating block using a water-based coupling gel and slight compressive pressure between the blocks enables a two-foci element with minimal energy loss at the block interface (Figs.~\ref{fig:fig4} and~\ref{fig:SI_Fig_6}). Moreover, these phase-transforming elements can be flexibly rearranged and replaced, in a manner common in optical assemblies but not previously demonstrated for MHz-frequency ultrasound.

\subsection*{Discussion}
Here, we have introduced a metamaterial architecture for MHz-frequency ultrasound based on microstructured glass. By patterning subwavelength laser-induced microcracks in regular grids, it is possible to simultaneously lower the effective speed of sound and raise the effective density of the glass by up to 20\%. In this work, the 20\% contrast limit was set by the minimum crack spacing, which we determined empirically as the smallest spacing achievable at our laser intensity that did not introduce unwanted fractures. Similarly, we left a safety margin of 4 mm between the metamaterial patterns and the glass-air interface. While these margins ultimately limit how small we can make a $2\pi$ phase modulation pattern with the current parameters (20 mm device at 2.25 MHz), optimizing both the glass material and processing parameters will allow for more compact devices and higher contrasts. For instance, the maximum achievable index contrast ($\approx$20\%) could be further increased by strengthening crack–wave interactions through higher crack density or larger crack size (e.g., via increased laser intensity). To avoid shattering, such changes would need to be optimized empirically alongside glass formulation, tempering procedures, and laser pulse duration. Building upon recent advances in femtosecond laser machining, it is reasonable that sub-micrometer cracks can be produced with
higher density and therefore that device size could be reduced by at least two orders of magnitude for higher-frequency ultrasound applications~\cite{Glezer1997Aug,Glezer1997May,Kruger1998May,Schrodter2025Jul,Gao2024Dec}. It is also possible to scale our approach toward larger apertures. Transducers with an aperture twice
as big are commonly commercially available and are directly applicable to larger glass geometries. To reach higher frequencies up to 10 MHz, one needs at least to shorten the microcrack spacing $s_i$, according to the scale of the problem $s_i/\lambda$, which introduces the manufacturing challenges already mentioned. For higher frequencies, alternative two-dimensional (2D) manufacturing approaches based on photolithography may be more suitable (see Tables~\ref{tab:table_literature_3d} and~\ref{tab:table_literature_2d}).

Our metamaterial allows for manipulations of incident ultrasonic wavefronts with minimal reflections at the metamaterial interface. The metamaterial's effective properties emerge from collective interactions between the anisotropic cracks, which we capture theoretically using the multiple-scattering \textit{T}-matrix formalism. Numerical predictions from our \textit{T}-matrix framework demonstrate excellent agreement with experimental results and capture the anisotropic behavior of the metamaterial arising from the microcrack response. The numerical framework provides a powerful tool for further exploring the design space of such metamaterials, including how microstructure patterns can be designed to achieve specific effective properties.

At MHz frequencies, few metamaterial architectures have been developed, and many traditional architectures that have good performance at lower frequencies do not scale well because of increased losses and manufacturing limitations. The approach presented here overcomes these limitations by leveraging a natively 3D microfabrication technique to directly modify a low-loss substrate, namely glass. In comparison with other common MHz-frequency metamaterials, such as elastomer composites or steel-rod phononic crystals~\cite{martinez_ultrasound_2025}, this approach offers numerous advantages. Microstructured glass can be used to create truly 3D metamaterials~\cite{kadic_3d_2019} from a mechanically and thermally stable material. By embedding the metamaterial in glass blocks with flat faces, metamaterial devices can easily be index-matched to ambient conditions using standard index-matching layers. Moreover, laser refracto-vibrometry can be used to directly interrogate the ultrasonic fields inside the material to help drive the development and optimization of metamaterial devices. In contrast with other microstructured metamaterials based on 3D-printed polymers~\cite{sun_tailored_2024}, the metamaterial introduced here exhibits extremely low losses at MHz frequencies and remains stable up to high intensities (at least 20 W/cm$^2$). Together, these advantages provide opportunities for the development of metamaterial-based ultrasonic devices at MHz frequencies.

One of the promising directions in which the metamaterial can provide enhanced capabilities is complex 3D wavefront shaping. Because of the manufacturing method  and the deep-subwavelength sizes of the fundamental scatterers, the metamaterial can easily be spatially patterned in 3D for complex wavefront engineering. We demonstrate this capability by creating a library of patterned metamaterial blocks that perform diverse wavefront-shaping functionalities, including focusing, diverting energy into grating lobes, and patterning sound using an acoustic hologram. We are not aware of any metamaterial
system to date that has been used for such general wavefront manipulations at MHz frequencies. Moreover, the metamaterial architecture is highly modular, and the fundamental blocks can be combined in different ways to produce larger wavefront-shaping assemblies. Such assemblies are commonplace in optics, where they provide the flexibility to define intricate and diverse optical experiments, yet such acoustic elements have not previously been developed for ultrasound. Microstructured glass can
therefore provide substantial opportunities to develop customized wavefront-shaping systems, particularly for high-power applications, including contactless ultrasonic power transfer~\cite{bakhtiari-nejad_acoustic_2018}, therapeutic focused ultrasound for ablation and hyperthermia~\cite{cengiz_roadmap_2025}, and manufacturing~\cite{derayatifar_holographic_2024}.

Beyond conventional applications of wavefront shaping, microstructured glass offers unique benefits that could facilitate easier and more widespread realization of MHz-frequency ultrasonic devices. In particular, the 3D patternability, high speed of sound, and low losses together make the metamaterial particularly appealing for potential applications in ultrasonic signal processing and computing~\cite{hughes_wave_2019}, especially in opening the door to volumetric ultrasonic neural processing.

\subsection*{Materials and Methods}

\subsubsection*{Characterization of the bulk glass material}
Glass samples were ordered from the company \textit{M+B Lasertechnik GmbH}. Most ordered glass blocks were cubes with an edge length of 50 mm. Three control cubes, i.e., without any engraving, were studied first. Their density was determined by weighing them and measuring their volume. The longitudinal speed of sound was measured with an Olympus transducer with a diameter of 1.5 inches (38.1 mm) and with a center frequency of 2.25 MHz. A refracto-vibrometry technique was used to measure the time-of-flight. Additional details on the refracto-vibrometry measurement, as well as a schematic (Fig.~\ref{fig:SI_Fig_7}A) and a photograph (Fig.~\ref{fig:SI_Fig_8}A) of the setup, are provided in Ref.~\cite{methods}. From a five-cycle burst at 2.25 MHz, times-of-flight $t_{1}$ and $t_{2}$ were measured at one position but for several time subsets corresponding to different reflections of the burst from the bottom face of the cube (Figs.~\ref{fig:SI_Fig_7}B and C). The longitudinal speed of sound was then calculated as 
\begin{equation}
	v_0 = \frac{2L}{t_{2} - t_{1}}\,.
	\label{eq:eq1} 
\end{equation}
Multiple points along the propagation direction were measured using the same protocol to obtain independent data points and assess the variability of the speed of sound measurements. Refracto-vibrometry provides a 2D projection of the 3D acoustic field and therefore does not directly capture the full spatial pressure distribution. We used a standard pulse-echo measurement to validate the robustness of the refracto-vibrometry approach for speed of sound measurement. The same transducer was connected to a pulser-receiver along with an oscilloscope (Fig.~\ref{fig:SI_Fig_7}D). By measuring the times-of-flight $t_{1}$ and $t_{2}$ of two consecutive amplitude peaks of the obtained time signal, i.e., two consecutive reflections, we calculated the speed of sound using Eq.~\eqref{eq:eq1}. Several consecutive peaks were used to obtain independent data points. Both techniques (refracto-vibrometry and pulse echo measurements) gave a 0.1\,\% difference (relative ratio).

Finally, we measured the attenuation of the glass material at several MHz frequencies. For this, we used a bigger block of size $70 \times 70 \times 250$ mm to increase the propagation distance between consecutive reflections. We used the same pulse-echo setup (Fig.~\ref{fig:SI_Fig_5}A) with transducers whose center frequencies are 2.25, 5, and 10 MHz. The detected absolute maximum magnitude for each reflection was plotted as a function of time which can be converted as the propagation distance, knowing the speed of sound (Fig.~\ref{fig:SI_Fig_5}B). The slope of a linear fit at each frequency provided the attenuation. Another linear fit was used to derive the attenuation per MHz.

\subsubsection*{Characterization of engraved glass area as an anisotropic effective medium}
After characterizing the bulk glass material, we investigated the properties of the engraved area. To this end, we considered a sample with one engraved metavoxel of size $4 \times 4 \times 31$ mm consisting of a regular 3D lattice of microcracks (Fig.~\ref{fig:fig2}A) and compared it to a control block. The metavoxel was elongated along one direction  because the refracto-vibrometry measurement is a 2D projection of a 3D field. The metavoxel length along the optical path of the vibrometer's laser had to be sufficiently large to yield a meaningful measurement. Apart from this constraint for the characterization with refracto-vibrometry, the metavoxel can have any size. A 3D-printed base allowed for the placement of both glass cubes at the same position (Fig.~\ref{fig:SI_Fig_8}A). Using refracto-vibrometry, we measured the time-of-flight between two scanning points for both blocks. Considering the engraved area as an effective medium, we calculated the effective speed of sound and the effective density for each propagation direction from the time-of-flight data (Figs.~\ref{fig:fig2}C and~\ref{fig:SI_Fig_9}). For more details on the calculation of the effective speed of sound and density, see Ref.\cite{methods}. 
Finally, we measured the attenuation inside the engraved area. Three large cylinders were separately engraved along the $x$, $y$, and $z$-axes in three different glass cubes. We used the same approach as mentioned above to measure the attenuation for each of them at 2.25 MHz.

\subsubsection*{\textit{T}-matrix simulation}
To compute the scattered pressure field of microcrack patterns and also evaluate their effective parameters, we used a \textit{T}-matrix-based approach that has been implemented in the open-source Python package \textit{acoustotreams}, v.0.2.5. This full-wave simulation approach considers the microcracks as anisotropic ellipsoids in the quasistatic approximation and computes their linear acoustic response by expanding incident and scattered pressure fields in scalar multipole waves, including only the monopole and the dipole. The \textit{T}-matrix, which can analytically account for collective interactions between multiple microcracks, directly maps the incident coefficients to the scattered ones, enabling efficient and accurate predictions for multiple patterned microcracks (Fig.~\ref{fig:fig2}D). For more details on the \textit{T}-matrix approach, see Ref.~\cite{methods}.

\subsubsection*{Design and measurement of a few metamaterials}
After experimentally and theoretically characterizing an engraved voxel as an anisotropic effective medium, we designed several standard elements to shape ultrasonic wavefronts inside the glass block, such as a biconvex converging lens, a 3D cylinder acting as an acoustic waveguide, and a 1D grating (Fig.~\ref{fig:fig3}B--D). For every 3D structure to be engraved, we generated a 3D cloud of points spaced by [90, 90, 270] \textmu m, so that the resulting pattern has the corresponding shape of a desired structure. The list of millions of $(x,y,z)$ coordinates was then exported as a text file and sent to the company for laser engraving. A pulsed laser operating at a wavelength of 532~nm was directed along the $z$-axis at each point to generate a microcrack. Additional details on the laser-engraving step are provided in Ref.~\cite{methods}. To characterize the resulting pressure fields, we used the same laser refracto-vibrometry again. We then obtained the 2D time-dependent acoustic signal measured by the vibrometer. We also computed the Fast Fourier Transform (FFT) of the measured signals using a time window that encompasses the acoustic burst propagation within the field of view. The amplitude and the phase of the FFT were then plotted to visualize the field.

\subsubsection*{Design and measurement of acoustic holograms}
\begin{figure}
	\centering
	\includegraphics[width=0.8\textwidth]{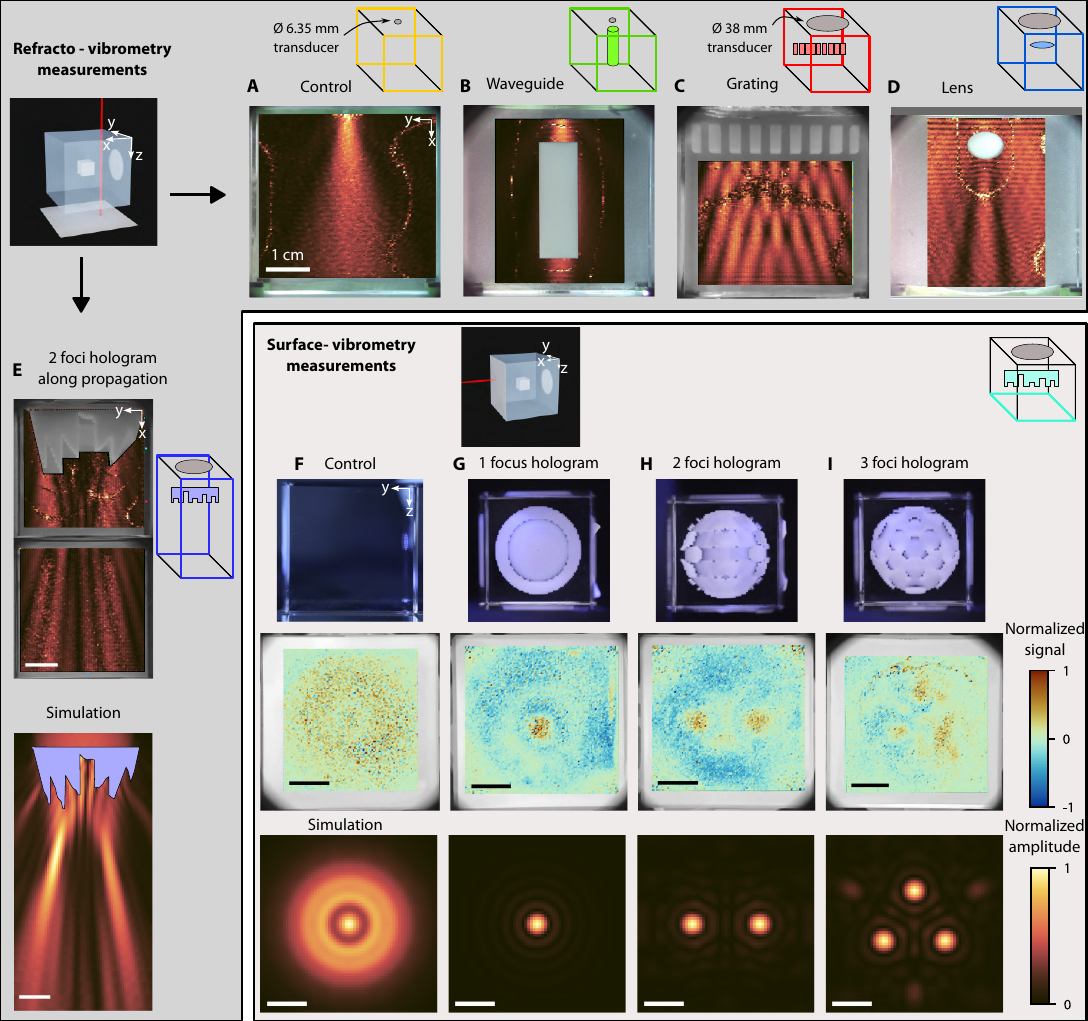}
	\caption{\textbf{Element library.}
    Different blocks were made to control the wavefront inside the glass. An ultrasound signal was transmitted along the $x$-direction. \textbf{A} to \textbf{E} were measured with a refracto-vibrometry technique, revealing: (\textbf{A}) a control block without any engraving, (\textbf{B}) a waveguide,  (\textbf{C}) a grating, (\textbf{D}) a biconvex converging lens, and (\textbf{E}) a hologram with two foci. For (\textbf{E}), a simulation is also provided for comparison. \textbf{F} to \textbf{I} were measured with a standard surface vibrometry technique, revealing: (\textbf{F}) a control block, (\textbf{G}) a one-focus hologram, (\textbf{H}) a two-foci hologram, and (\textbf{I}) a three-foci hologram whose image plane coincides with a face of the cube opposite the transducer. For \textbf{F} to \textbf{I}, top-down photographs of the samples are provided (top), as well as the vibrometry time signal (middle), and simulations (bottom) for comparison.
    For all the blocks, a schematic of the setup is depicted. The cube's face, which was measured, is highlighted in color. For the refracto-vibrometry, a 3D sound field is reduced to a 2D projection. Two opposite faces are highlighted to depict the measurement direction. All scale bars indicate 1 cm.}
	\label{fig:fig3}
\end{figure}

\begin{figure}
	\centering
	\includegraphics[width=1\textwidth]{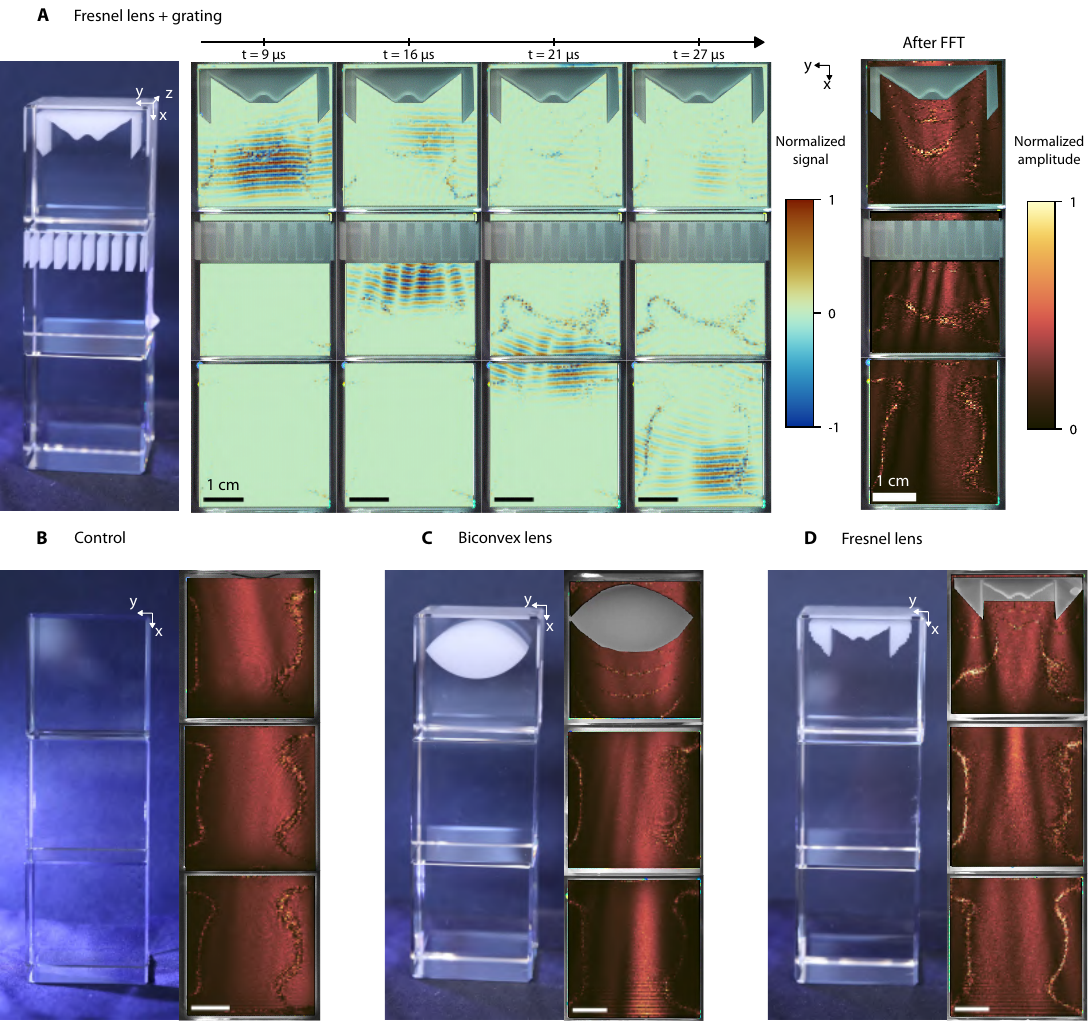}
	\caption{\textbf{Stacking of blocks for modularity.}
		 (\textbf{A}) Photograph of the assembly of a converging Fresnel lens and a grating. The refracto-vibrometry signal shows the propagation of the field with limited reflection between the blocks. After the Fast Fourier Transform, the converging effect as well as the grating are visible. (\textbf{B})-(\textbf{D}) Photograph and FFT of the refracto-vibrometry signal for (\textbf{B}) a control assembly (without any engraving), (\textbf{C}) a biconvex converging lens, and (\textbf{D}) a converging Fresnel lens. All scale bars indicate 1 cm.}
	\label{fig:fig4}
\end{figure}

Phase plates were designed and engraved inside the glass blocks (Fig.~\ref{fig:fig3}). The phase plates were positioned 4~mm from the face in contact with the transducer and had approximately the same diameter as the transducer aperture. One hologram was designed to generate two simultaneous focal spots at different depths along the propagation axis (Fig.~\ref{fig:fig3}E). The same refracto-vibrometry technique was used to measure the field. In this case, a longer ultrasound burst consisting of 10 cycles was transmitted instead of the one with five cycles. Several additional phase plates were designed such that the image plane coincided with the cube face opposite the transducer (Fig.~\ref{fig:fig3}G--I). This time, conventional surface vibrometry technique was employed instead of refracto-vibrometry. The cube face located at the image plane was coated with a white spray to increase optical reflectivity for the vibrometer measurements. The same vibrometry parameters were used, and a 10-cycle ultrasound burst was transmitted. A photograph of the setup is shown in Fig.~\ref{fig:SI_Fig_8}B. As a control sample, a glass block with no engraving was also characterized (Fig.~\ref{fig:fig3}F).
To produce the desired wavefront at the image plane, we employed the Iterative Angular Spectrum Approach (IASA) to compute the relative phase shifts across the phase-plate pixels, following the method described in Ref.~\cite{melde_holograms_2016}. Because the phase plate was located 4~mm from the transducer, the acoustic field--assumed to possess a uniform phase at the transducer surface--was first numerically propagated to the plane of the phase plate. The required phase distribution for the plate design was then obtained by iteratively propagating the field between the image plane and the phase-plate plane while applying amplitude constraints in both planes at each iteration. The resulting phase distribution was then converted into a physical pixel height using the equation $h = \frac{\Delta\phi}{2\pi f_0} \times \frac{v_0v^{\rm eff}_x}{(v_0 - v^{\rm eff}_x)}$, where $\Delta\phi$ denotes the phase shift for an individual pixel. Finally, the final phase-plate geometry was converted into a 3D point cloud for laser engraving. For the hologram generating two focal spots along the propagation direction, a simplified design procedure was used: two spherical waves were numerically back-propagated from the desired focal positions, and the resulting fields were superposed. Its phase was then used to similarly design the corresponding phase plate.

\subsubsection*{Stacking of samples for modularity}
Different combinations of glass blocks were stacked to form modular assemblies (Fig.~\ref{fig:fig4}). A simple scaffold was 3D printed to align the stacked glass blocks. A thin layer of water was used for coupling between the blocks, while two screws applied light compressive pressure to reduce reflections of the acoustic signal at the interfaces. Three scans were then performed with the vibrometer to characterize the entire assembly (Fig.~\ref{fig:SI_Fig_6}).

\subsubsection*{Statistics}
For all line plots, the distribution is depicted as follows: mean values are shown by markers, and whiskers represent $\pm$ one standard deviation. All values are presented as ``mean value $\pm$ standard deviation'' unless otherwise specified.

\begin{table}[h!]
	\centering
	\caption{\textbf{Attenuation of different materials}
		Attenuation for water, soft tissue~\cite{shankar_potential_2011}, BK7 glass~\cite{ono_comprehensive_2020}, steel~\cite{ono_comprehensive_2020}, silicone~\cite{itsumi_low_2009}, and VeroClear~\cite{bakaric_measurement_2021}, a photopolymer traditionally used for 3D printed phase plates.}
	\label{tab:tab1} 
    \begin{tabular}{l|l|l|l|l|l|l|}
    \cline{2-7}
    & Water & BK7 glass & A36 steel & Soft tissue  & VeroClear & Silicone \\ \hline
    \multicolumn{1}{|l|}{\begin{tabular}[c]{@{}l@{}}Attenuation\\ (dB\,MHz$^{-1}$\,cm$^{-1}$) \end{tabular}} & 0.0022 & 0.05 & 0.124 & 0.75 & 3.7 & 4  \\ \hline
    \end{tabular}
\end{table}

\begin{table}[h!]
\centering
\caption{\textbf{Effective properties of the engraved area (metamaterial)}
Density, longitudinal speed of sound, and attenuation of the bulk medium (BK7 glass) and of the engraved area, which is considered as an effective medium.}
\label{tab:tab2} 
\resizebox{\linewidth}{!}{
\begin{tabular}{c|c|ccc|}
\cline{2-5}
                                                  & Bulk medium                                  & \multicolumn{3}{c|}{Effective medium} \\ \cline{3-5} 
                                                  &                                              & \multicolumn{1}{c|}{$x$}                   & \multicolumn{1}{c|}{$y$}                   & {$z$} \\ \hline
\multicolumn{1}{|c|}{Density}                     & $\rho_0 = 2525\pm8$ kg\,m$^{-3}$              & \multicolumn{1}{c|}{$\left(1.21\pm0.10\right)\rho_0$} & \multicolumn{1}{c|}{$\left(1.03\pm0.09\right)\rho_0$} & $\left(1.02\pm0.03\right)\rho_0$ \\ \hline
\multicolumn{1}{|c|}{Longitudinal speed of sound} & $v_{0} = 5816\pm4$ m\,s$^{-1}$                & \multicolumn{1}{c|}{$\left(0.83\pm0.07\right)v_{0}$}  & \multicolumn{1}{c|}{$\left(0.98\pm0.08\right) v_{0}$} & $\left(0.98\pm0.03 \right)v_{0}$ \\ \hline
\multicolumn{1}{|c|}{Attenuation}                 & $\alpha = 3.3\pm0.4$ dB\,MHz$^{-1}$\,m$^{-1}$ & \multicolumn{1}{c|}{$\left(1.1\pm0.1\right)\alpha$}   & \multicolumn{1}{c|}{$\left(1.3\pm0.1\right)\alpha$}   & $\left(1.2 \pm 0.1\right) \alpha$ \\ \hline
\end{tabular}
}
\end{table}

\clearpage 
%
\bibliography{low_attenuation_MHz_metamaterial}
\bibliographystyle{sciencemag}

%
%
%
%
%
%

\paragraph*{Acknowledgments:}
We thank Dr. Jochen Schell from the Polytec company for technical support (PSV-500) and for helpful comments regarding refracto-vibrometry measurements. We also thank Mr. Holger Hartmann from M+B Lasertechnik GmbH for his expertise and advice concerning the laser engraving of glass samples. We are grateful to Dr. Kai Melde for helpful discussions and for providing transducers, and to Setthibhak Suthithanakom for his help with the design of the 3D-printed holder. The authors gratefully acknowledge the data storage service SDS@hd supported by the Ministry of Science, Research, and the Arts Baden-Württemberg (MWK) and the German Research Foundation (DFG) through grant INST 35/1503-1 FUGG.

\paragraph*{Funding:}
This work was supported by the Deutsche Forschungsgemeinschaft (DFG, German Research Foundation) under Germany’s Excellence Strategy via the Excellence Cluster 3D Matter Made to Order (EXC-2082/2, Grant No. 390761711 to P.F and C.R.) and from the Carl Zeiss Foundation (Grant CZF-Focus@HEiKA to P.F and C.R).

\paragraph*{Author contributions:}
	Conceptualization: OD, AGA, CR, PF \\
	Methodology: OD, NU, AGA, CR, PF \\
	Investigation: OD, NU, AGA, LG \\
	Visualization: OD, NU \\
    Resources: PF \\
	Supervision: AGA, CR, PF \\
	Writing—original draft: OD, NU \\
	Writing—review \& editing: OD, NU, AGA, LG, CR, PF 

\paragraph*{Competing interests:}
All authors declare they have no competing interests.

\paragraph*{Data, Code, and Materials Availability:}
All data and code needed to evaluate and reproduce the results in the paper are present in the paper and/or the Supplementary Materials. The experimental data generated in this study as well as the code and results of the T-matrix simulation have been deposited in the Zenodo database under accession code https://doi.org/10.5281/zenodo.19919740. Requests should be submitted to oscar.demeulenaere@mr.mpg.de


\subsection*{Supplementary materials}
Materials and Methods\\
Figs. S1 to S11\\
Tables S1 to S6\\
References \textit{(44-\arabic{enumiv})}\\ 


\newpage


\renewcommand{\thefigure}{S\arabic{figure}}
\renewcommand{\thetable}{S\arabic{table}}
\renewcommand{\theequation}{S\arabic{equation}}
\renewcommand{\thepage}{S\arabic{page}}
\setcounter{figure}{0}
\setcounter{table}{0}
\setcounter{equation}{0}
\setcounter{page}{1} 


\begin{center}
\section*{Supplementary Materials for\\ \scititle}

Oscar~Demeulenaere$^{\ast}$,
Nikita~Ustimenko,
Athanasios~Athanassiadis\\
Lovish~Gulati,
Carsten~Rockstuhl,
Peer~Fischer\\ 
\small$^\ast$Corresponding author. Email: oscar.demeulenaere@mr.mpg.de
\end{center}

\subsubsection*{This PDF file includes:}
Materials and Methods\\
Figures S1 to S11\\
Tables S1 to S6\\


\newpage


\subsection*{Materials and Methods}

\subsubsection*{Refracto-vibrometry measurements}
As explained in Ref.~\cite{zipser_5e-4_2007}, refracto-vibrometry can be used in transparent solids. With refracto-vibrometry, a 3D sound field is reduced to a 2D projection. To verify that this projection did not interfere with our time-of-flight measurements, we also measured the speed of sound in the bulk medium with a pulser-receiver as shown in Fig.~\ref{fig:SI_Fig_7}.

The refracto-vibrometry setup was composed of a Polytec Scanning Vibrometer PSV-500, a signal generator, an amplifier, and an ultrasound transducer. The acquisition parameters are described in Table~\ref{tab:table_s2}. We placed the glass cube inside a 3D printed base and placed a reflective surface behind it. A photograph of the setup is provided in Fig.~\ref{fig:SI_Fig_8}. We calibrated the position between the scan points from the distance from the edge of the glass cube (as recorded on camera). To focus the laser spot, we removed the glass cube and moved the reflective surface to the middle plane of the glass cube. The laser spot was then sharp inside the glass cube, but a bit bigger on the reflective screen, back to its original position behind the cube. The refracto-vibrometry time-domain signal for each scanning point was then exported as a .uff file. The raw data were imported and analyzed with the Fast Fourier Transform (FFT) using Python 3.12.

\subsubsection*{Measurement of the effective speed of sound and density with refracto-vibrometry}
To measure the effective speed of sound, we considered two samples: a control block and a block with an engraved voxel at the center (see Fig.~\ref{fig:SI_Fig_9}). Both blocks were positioned inside a 3D-printed base to ensure the positions of the scanning points are identical in both measurements. We then used an Olympus transducer with a diameter of 6.35 mm and a center frequency of $f_{0} = 2.25$ MHz. A five-cycle burst was transmitted at $f_{0}$. Using the refracto-vibrometry data, we measured the time-of-flight between two scanning points separated by a fixed distance $d$. We denoted these values as $t_{1}^{\rm C}$ and $t_{2}^{\rm C}$ for the control block, and $t_{1}^{\rm E}$ and $t_{2}^{\rm E}$ for the engraved block (Fig.~\ref{fig:SI_Fig_9}). Assuming the impedance matching condition, the effective speed of sound was then calculated as $v^{\rm eff}_i = \left(\frac{t_{2}^{\rm E} - t_{1}^{\rm E}~+~t_{2}^{\rm C} - t_{1}^{\rm C}}{h} + \frac{1}{v_0}\right)^{-1}$, where $\rho_0$ and $v_0$ are the density and speed of sound of the bulk medium, respectively, and $h$ is the height of the engraved voxel along the propagation direction of the ultrasound beam. To calculate the effective density, we used the equation $\rho^{\rm eff}_i = \frac{\rho_{0}v_{0}}{v^{\rm eff}_i}$, which follows from the impedance matching condition. For more details on this assumption, see Ref.\cite{methods}. This procedure was repeated for all three propagation directions ($ i = x , y, z$).

Since the 3D sound field is reduced to a 2D projection in refracto-vibrometry, the size of the engraved voxel along the vibrometry laser relative to the block size affects the measured signal and, consequently, the estimation of the effective parameters. Therefore, to determine the effective parameters correctly, we considered an engraved voxel that was sufficiently large along the refracto-vibrometry laser direction (Fig.~\ref{fig:SI_Fig_2}).

\subsubsection*{Measurement of the reflection coefficient of the metamaterial and correction to the effective density}

The impedance-matching condition implies negligible reflection. To validate this assumption, we measured the reflection coefficients $r_i$ of the metamaterial for each propagation direction $i$ and evaluated the corresponding corrections to the effective densities arising from the finite reflection.

For an incident Gaussian beam with pressure field in the frequency domain 
$p_{\rm i}(f_0, \mathbf{r}) = A(f_0, \mathbf{r})\mathrm{e}^{\mathrm{i}\Phi(f_0, \mathbf{r})}$, where $A(f_0, \mathbf{r})$ and $\Phi(f_0, \mathbf{r})$ denote the beam amplitude and phase, respectively, the reflected beam can be expressed as 
$p_{\rm r}(f_0, \mathbf{r}) = r_i(f_0)A(f_0, \mathbf{r})\mathrm{e}^{-\mathrm{i}\Phi(f_0, \mathbf{r})}$, where $ r_i(f_0)=|r_i(f_0)|\mathrm{e}^{\mathrm{i}\theta_{i}(f_0)}$ is the complex reflection coefficient at frequency $f_0$~\cite{Svelto2010}. The amplitude and phase of the reflection coefficient are therefore given by
\begin{align}
\label{eq:r_from_fields}
    |r_i(f_0)| = \left|\frac{p_{\rm r}(f_0, \mathbf{r})}{p_{\rm i}(f_0, \mathbf{r})} \right|\,, \quad
    \theta_i(f_0) = \arg\left[ p_{\rm r}(f_0, \mathbf{r})\right] + \arg\left[ p_{\rm i}(f_0, \mathbf{r})\right]\,,
\end{align}
respectively. 

The incident and reflected fields at 2.25~MHz were evaluated at several points above the metamaterial and substituted into Eq.~\eqref{eq:r_from_fields}. The resulting values of $r_i$ were then averaged. To convert the measured signals in the time domain to the frequency domain, we applied the FFT to the refracto-vibrometry data using two time windows of equal duration to isolate the incident and reflected beams.

Assuming that the coefficients $r_i$ correspond to the reflection coefficient of a slab system (glass--metamaterial--glass), we determined the effective reflection coefficient of the glass--metamaterial interface by finding a root to the following equation such that $\left| r^{\rm eff}_i\right|^2 \leq 1$: 
\begin{align}
\label{eq:r_eff}
    \left(r_i \mathrm{e}^{\mathrm{i} 2 \phi^{\rm eff}_i}\right) \left(r^{\rm eff}_i\right)^2  +  \left(1- \mathrm{e}^{\mathrm{i} 2 \phi^{\rm eff}_i}\right)r^{\rm eff}_i -r_i = 0\,,
\end{align}
where $\phi^{\rm eff}_i = 2 \pi f_0 h / v^{\rm eff}_i$ is the effective phase accumulation~\cite{LeBas2005Mar}. Here, $h$ is the height of the engraved voxel along the propagation direction of the ultrasound beam. 

From $r^{\rm eff}_i$, we calculated the correction to the effective density as 
\begin{align}
\label{eq:delta_rho}
    \delta \rho^{\rm eff}_i = \rho^{\rm eff}_i \left[\Re\left( \frac{1 + r^{\rm eff}_i}{1 - r^{\rm eff}_i}\right) - 1\right]\,.
\end{align} 
The values of the reflection coefficients and the density corrections are summarized in Table~\ref{tab:table_density_num}. Importantly, the condition $|\delta \rho^{\rm eff}_i| \ll \rho^{\rm eff}_i$ is satisfied, thereby supporting the initial assumption of the impedance matching.

\subsubsection*{Measurement of the effective speed of sound for different frequencies}
A setup similar to the one above was used to measure the  frequency dependence of the effective parameters of the engraved area. We used four different transducers whose center frequencies were 0.5, 1, 2.25, and 5 MHz. The control sample remained the same as above, while a waveguide-like arrangement of microcracks was used as the engraved sample. For this engraved sample, the size of the pattern of microcracks along the vibrometry laser direction is 10 mm. Figure~\ref{fig:SI_Fig_4} shows the effective speed of sound along the $x$-axis as a function of frequency evaluated in the time domain via time-of-flight measurements. From the spectrum, the operational bandwidth (full width at half maximum) is 3.1 MHz, corresponding to a fractional bandwidth of 87\%. 

\subsubsection*{\textit{T}-matrix method for simulating microcrack arrays}
Below we present a theoretical description of the acoustic response of the microcrack metamaterial sketched in Fig.~\ref{fig:SI_Fig_10}. We assume that the microcracks are identical and consider each microcrack as an anisotropic ellipsoid with semiaxes $a_x = 19.7$ $\mu$m, $a_y = 19.45$ $\mu$m, and $a_z = 105.5$ $\mu$m along the $x$-, $y$-, and $z$-axes, respectively, resulting in a microcrack volume of $V_{\rm e} = \frac{4\pi}{3}a_x a_y a_z$. We also assume that the principal axes of the microcrack material coincide with the Cartesian coordinate axes. Then, the microcrack has density components $\{ \rho_x, \rho_y, \rho_z\}$ and a bulk modulus $K$, while the isotropic background medium has the density $\rho_{0} = 2525$ kg\,m$^{-3}$, the speed of sound $v_{0} = 5816$ m\,s$^{-1}$, and the bulk modulus $K_{0} = \rho_{0}v_{0}^2 \approx 85.4$ GPa at frequency $f_0 = 2.25$ MHz (Table~\ref{tab:tab2}). The attenuation and shear waves in the glass medium are also neglected. Therefore, we can treat the bulk glass medium as a non-viscous fluid. 

Comparing the microcrack size ($2a_z \approx 0.2$ mm) and spacing ($s_z \approx 0.27$ mm) with the wavelength in the glass medium ($\lambda = v_0/f_0 = 2.58$ mm at $f_0 = 2.25$ MHz) results in the ratios $2a_z \approx \lambda/13$ and $s_z \approx \lambda/9$. Therefore, the considered metamaterial is a deeply subwavelength structure, which may be treated as an effective medium. Accordingly, the acoustic response of the microcracks can be described within the quasistatic approximation.

\noindent\textbf{Estimation of the microcrack material parameters through the effective medium theory}

To find $K$, we apply Wood's law~\cite{Wood1941,Torrent2006Dec}
\begin{align}
\label{eq_eff_K}
    \frac{1}{\widetilde{K}^{\rm eff}} = \frac{\eta}{ \widetilde{K}}+(1-\eta)\,,
\end{align}
where the bulk moduli have been normalized by $K_0$, and the fraction of the microcrack volume to the unit cell volume, i.e., the filling factor is $\eta \approx 0.077$. 

To find $\rho_i$, we adopt the approach developed for a diluted ($\eta \ll 1$) ensemble of electromagnetic ellipsoidally-shaped scatterers~\cite{Giordano2003May}. Let a uniform velocity field $\mathbf{u}_0$ be applied to the ellipsoidal microcrack. Then, the field inside the microcrack is also uniform, and its components are given by
$
    u_{ix} = \frac{u_{0x}}{\widetilde{\rho}_{x} + L_x \left(1 -\widetilde{\rho}_{x}  \right)}\,,$ $ u_{iy} = \frac{u_{0y}}{\widetilde{\rho}_{y} + L_y \left(1 -\widetilde{\rho}_{y}  \right)}\,, $ and $u_{iz} = \frac{u_{0z}}{\widetilde{\rho}_{z} + L_z \left(1 -\widetilde{\rho}_{z}  \right)}\,,
$
where the densities have been normalized by $\rho_0$, and the depolarization factors of the considered ellipsoidal microcrack have been computed as $L_x = 0.472$, $L_y = 0.478$, and $L_z = 0.05$~\cite{Smagin2024Dec}. Next, since the microcrack array is elongated along the laser direction, the volume-averaged effective field $\mathbf{u}^{\rm eff} = \eta \mathbf{u}_i + (1 - \eta) \mathbf{u}_0$ can be approximated by the field generated by a large ellipsoid of the same shape, so that
$
    u_{x}^{\rm eff} = \frac{u_{0x}}{\widetilde{\rho}_{x}^{\rm eff} + L_x \left(1 -\widetilde{\rho}_{x}^{\rm eff}  \right)}\,,$ $u_{y}^{\rm eff} = \frac{u_{0y}}{\widetilde{\rho}_{y}^{\rm eff} + L_y \left(1 -\widetilde{\rho}_{y}^{\rm eff}  \right)}\,,$ and $u_{z}^{\rm eff} = \frac{u_{0z}}{\widetilde{\rho}_{z}^{\rm eff} + L_z \left(1 -\widetilde{\rho}_{z}^{\rm eff}  \right)}\,.
$
Hence, the components of the effective density tensor can be calculated as follows 
\begin{align}
\label{eq_eff_rho}
 \left[\widetilde{\rho}^{\rm eff}_i\right]^{-1} = 1 + \frac{\eta\left(\left[\widetilde{\rho}_i\right]^{-1}-1 \right)}{1 + (1-\eta)L_i\left(\left[\widetilde{\rho}_i\right]^{-1}-1 \right)}\,.
\end{align}

As the effective densities of the microcrack array, we take the following values $\widetilde{\rho}_x^{\rm eff} = 1.21$, and $\widetilde{\rho}_y^{\rm eff} = \widetilde{\rho}_z^{\rm eff} = 1$ (Table~\ref{tab:tab2}). To calculate the effective bulk modulus, we take the normalized speed of sound $\widetilde{v}^{\rm eff}_x \equiv v^{\rm eff}_x/v_0 = 0.83$ (Table~\ref{tab:tab2}) and calculate $\widetilde{K}^{\rm eff} = \widetilde{\rho}^{\rm eff}_x(\widetilde{v}^{\rm eff}_x)^2 \approx 0.83$ corresponding to the $x$-direction ($\widetilde{K}^{\rm eff} \equiv 1$ for the $y$- and $z$-axes). If we plug $\widetilde{K}^{\rm eff}$ and $\widetilde{\rho}^{\rm eff}_i$ into Eqs.~\eqref{eq_eff_K} and~\eqref{eq_eff_rho}, respectively, we obtain the normalized material parameters of the microcrack $\widetilde{K} = 0.27$, $\widetilde{\rho}_x = -7.80$, and $\widetilde{\rho}_y = \widetilde{\rho}_z = 1.00$. To calculate the normalized compressibility of the microcrack, we average the inverse bulk modulus, so that $\widetilde{\beta} = \left(\widetilde{K}^{-1} + 2\right)/3 \approx 1.90$, where $\beta_0 = K_0^{-1}$.

\noindent\textbf{Response of a microcrack in the quasistatic approximation}

As written above, we simulate the microcrack response in the quasistatic approximation, when the monopole and dipole polarizabilities of an ellipsoidally-shaped acoustic scatterer read as~\cite{Smagin2024Dec}
\begin{align}
    \label{eq_alpha_ellipsoid}
    \left(\alpha_{\rm M} \right)^{-1} = \left(\alpha_{\rm M}^{\rm st}\right)^{-1} - \frac{\mathrm{i}k^3}{4\pi}\, , \quad \text{and} \quad \left(\boldsymbol{\alpha}_{\rm D} \right)^{-1} = \left(\boldsymbol{\alpha}_{\rm D}^{\rm st}\right)^{-1} - \frac{\mathrm{i}k^3}{12\pi}\mathbf{I}_3\,,
\end{align}
where $k = 2\pi/\lambda$, and $\mathbf{I}_n$ is the $n \times n$ identity matrix. The static monopole and dipole polarizabilities are given by  
\begin{align}
    \alpha_{\rm M}^{\rm st} = V_{\rm e} \left(\widetilde{\beta} - 1 \right)\,, \quad \text{and} \quad \left(\alpha_{\rm D}^{\rm st} \right)_{i} = V_{\rm e} \frac{\widetilde{\rho}_{i} - 1}{\widetilde{\rho}_{i} + L_i \left(1 -\widetilde{\rho}_{i}  \right)}\,, \quad i = x,y,z\,.
\end{align}
The polarizabilities in Eq.~(\ref{eq_alpha_ellipsoid}) determine the acoustic \textit{T}-matrix of an isolated microcrack in the monopole--dipole approximation at frequency $f_0$ as~\cite{Toftul2019Oct}
\begin{align}
\label{eq_Tmatrix}
    \mathbf{T} =  \begin{pmatrix}
        \mathrm{T}_{\mathrm{M}} & \mathbf{0} \\
        \mathbf{0}& \mathbf{T}_{\mathrm{D}}
    \end{pmatrix}\,
    =\frac{\mathrm{i}k^3}{12\pi}
    \begin{pmatrix}
        3\alpha_{\rm M} & \mathbf{0} \\
        \mathbf{0}& \mathbf{F}^{-1}_{\rm 3D} \boldsymbol{\alpha}_{\rm D} \mathbf{F}_{\rm 3D}
    \end{pmatrix}\,,
\end{align}
where matrices $\mathbf{F}_{\rm 3D}$ and $\mathbf{F}^{-1}_{\rm 3D}$ convert the Cartesian components of the dipole moment into the spherical-wave basis and vice versa, respectively,~\cite{Mun2020May}
\begin{align}
    \mathbf{F}_{\rm 3D} = \frac{1}{\sqrt{2}}\begin{pmatrix}
        1& 0& -1\\
        -\mathrm{i}& 0& -\mathrm{i} \\
        0& \sqrt{2}& 0
    \end{pmatrix}\,,
    \quad \text{and} \quad
    \mathbf{F}^{-1}_{\rm 3D} = \mathbf{F}^{\dagger}_{\rm 3D} = \frac{1}{\sqrt{2}}\begin{pmatrix}
        1& \mathrm{i}& 0\\
        0& 0& \sqrt{2} \\
        -1& \mathrm{i}& 0
    \end{pmatrix}\,,
\end{align}
where $\dagger$ denotes the Hermitian conjugate.

\noindent\textbf{Multiple-scattering formalism}

As the \textit{T}-matrix in Eq.~\eqref{eq_Tmatrix} is known, we can consider a finite array of $N \equiv N_x \times N_y \times N_z$ microcracks with spacings $s_x = 90$ $\mu$m, $s_y = 90$ $\mu$m, and $s_y = 270$ $\mu$m; hence, the array size along the axis $i=x,y,z$ is $d_i = (N_i - 1)s_i +2a_i$. In the following, we consider two coordinate systems: $(x,y,z)$ is associated with the principal axes of the microcrack, while $(x',y',z')$ corresponds to a permutation of $(x,y,z)$. In our model, an incident wave always propagates along the $(-x')$-direction. In Fig.~\ref{fig:SI_Fig_2}, $d \equiv d_{z'} \propto N_{z'}$ varies, whereas $N_{x'} = N_{y'} = 45$. In Fig.~\ref{fig:SI_Fig_9}, the thickness of the array along the $x'$-axis is denoted as $h \equiv d_{x'}$.

Within the \textit{T}-matrix method, collective interactions between scatterers can be considered via the analytically known translation coefficients of spherical waves~\cite{Stein1961}. Furthermore, we can model the acoustic response of an array that has a finite number of scatterers along all directions~\cite{Ustimenko2025Apr}. However, for a relatively high $N$, it becomes computationally demanding, if not impossible, since we need to invert a matrix of size $4N$ (under the monopole--dipole approximation). 

Therefore, to simplify and speed up computations, we assume periodic boundary conditions along the refracto-vibrometry direction ($z'$-axis) with a lattice constant of $s_{z'}$, i.e., $N_{z'} \to \infty$. In this case, the 3D scattering problem becomes a 2D one, and the interactions between an infinite number of microcracks can be described by so-called lattice sums [see Eq.~(D6) in Ref.~\cite{Ustimenko2025Oct}], which can be computed fast in \textit{acoustotreams}. Then, only the following $4 \times 4$ matrix must be inverted to obtain the \textit{T}-matrix of a microcrack chain in the spherical-wave basis under the monopole-dipole approximation
\begin{align}
\label{eq_Tchain}
    \mathbf{T}_{\rm chain} = \left(\mathbf{I}_4 - \mathbf{\Sigma} \mathbf{T} \right)^{-1} \mathbf{T}\,,
\end{align}
where $\mathbf{\Sigma} = 2 \sqrt{\pi} D_{3, 0,0} \mathbf{I}_4 + \sqrt{5\pi}D_{3, 2,0} \boldsymbol{\Xi}$, $\boldsymbol{\Xi} = \mathrm{diag}\left[0; 2/5; -1; 2/5\right]$, and the sums $D_{3,\ell,m}$ are defined by Eq.~(1) in Ref.~\cite{Beutel2023Jan} for 1D lattices, with $\mathbf{k}_{\parallel} = \mathbf{0}$ (normal incidence) and $\mathbf{r} = \mathbf{0}$. 

Next, we transform the matrix in Eq.~(\ref{eq_Tchain}) in the spherical-wave basis into the following $3 \times 3$ matrix in the cylindrical-wave basis using the same approach as in Ref.~\cite{Beutel2024Apr} but for the scalar waves. For the lattice constant $s_{z'} < \lambda$ and normal incidence, one can show that the transformation is given by
\begin{align}
\label{eq_Tchain_cw}
    \boldsymbol{\mathcal{T}}_{\rm chain} = \frac{\lambda}{2s_{z'}} \mathbf{U}\mathbf{T}_{\rm chain} \mathbf{V}\,,
\end{align}
where
\begin{align}
    \mathbf{U} = \frac{1}{\sqrt{2}}
    \begin{pmatrix}
        0& -\sqrt{3}& 0 & 0 \\
        \sqrt{2}& 0 & 0 & 0 \\
        0& 0& 0& -\sqrt{3}
    \end{pmatrix}\,, 
    \quad \text{and} \quad 
     \mathbf{V} = \frac{1}{\sqrt{2}}
    \begin{pmatrix}
        0& \sqrt{2}& 0 \\
        -\sqrt{3}& 0& 0 \\
        0& 0 & 0 \\
        0& 0& -\sqrt{3}
    \end{pmatrix}\,. 
\end{align}

Now, we can consider a finite array of $N_{x'} \times N_{y'}$ microcrack chains infinitely extended along the $z'$-axis. Let the coordinates of the chains in the $z'=0$ plane be $\mathbf{r}'_{lj} = (x'_l,y'_j,0)$ with $x'_l = \left( l - \frac{N_{x'} + 1}{2}\right)s_{x'}$ and $y'_j = \left( j - \frac{N_{y'} + 1}{2}\right)s_{y'}$, where $1 \leq l \leq N_{x'}$ and $1 \leq j \leq N_{y'}$. $N_{x'}$ and $N_{y'}$ are odd natural numbers. As an incident field, we consider a Gaussian beam integrated along the $z'$-axis from $z' = -L$ to $z' = L$, where $L$ is the size of the glass block. Thus, the incident field is 
\begin{align}
\label{eq_pressure_inc}
    p_{\rm inc}(x',y') \approx p_0(x',y') \mathrm{e}^{-\mathrm{i} k x'}\,,
\end{align}
where $p_0(x',y') = \frac{\sqrt{\pi} W_{x'}}{L} \mathrm{erf}\left( \frac{L}{2W_{x'}}\right)\frac{w_{x'_0}}{w_{x'}}\mathrm{exp}\left(-\frac{y^{'2}}{W_{x'}^2} + \mathrm{i} \psi_{x'} \right)\,$ accounts for the averaging effect along the vibrometer laser direction ($z'$), $\mathrm{erf}(...)$ is the error function, the complex-valued beam width is $W_{x'} = w_{x'}\left(1 + \mathrm{i} \frac{k w_{x'}^2}{2 R_{x'}}\right)^{-1/2}\,,$ the real-valued beam width is $ w_{x'} = w_{x'_0} \sqrt{1 + \frac{(x' - x'_0)^2}{l_R^{2}}}\,,$ the radius of curvature is $R_{x'} = (x' - x'_0) \left[ 1 + \frac{l_R^{2}}{(x' - x'_0)^2}\right]\,,$ the Gouy phase is $\psi_{x'} = \arctan\left[\frac{x' - x'_0}{l_R}\right]\,,$
and the Rayleigh length is $l_R = \pi w_0^2 / \lambda$ for waist radius $w_{x'_0} = 3.175$ mm at the top face of the glass block $x'_0 = L/2$~\cite{Svelto2010}. 

We can then write the total pressure field using the following approximation
\begin{align}
\label{eq_pressure}
    \boxed{p(x',y') \approx p_0(x',y') \times \left[ \mathrm{e}^{-\mathrm{i} k x'} + \xi\left(d_{z'}\right) p_{\rm sca}(x',y') \right]}\,, 
\end{align}
where the prefactor $\xi\left(d_{z'}\right) = \min \left\{1, \frac{d{z'}}{\sqrt{\pi}\left|W_{x' = (L-d_{x'})/2}\mathrm{erf}\left(\frac{L}{2W_{x' = 0}}\right)\right|}\right\}$ accounts for the finite thickness of the microcrack array along the $z'$-axis and for the finite radius of the beam waist at the top face of the array. The time dependence of the field in Eq.~\eqref{eq_pressure} is monochromatic, i.e., $\mathrm{e}^{-\mathrm{i}2\pi f_0 t}$. 

To find the scattered field $p_{\rm sca}(x',y')$, we first expand the plane-wave part and the scattered field in Eq.~\eqref{eq_pressure} in regular and singular cylindrical waves with the Bessel and first-kind Hankel functions, respectively,
\begin{align}
    \mathrm{e}^{-\mathrm{i} k x'} =  \sum_{l,j,m} \mathcal{B}_{i,j,m} \underbrace{J_m(k|\mathbf{R}'_{lj}|) \mathrm{e}^{\mathrm{i} m \phi_{\mathbf{R}'_{lj}}}}_{\mathrm{regular \ CWs}}\,, \quad p_{\rm sca}(x',y') = \sum_{l,j,m} \mathcal{A}_{l,j,m} \underbrace{H_m^{(1)}(k|\mathbf{R}'_{lj}|) \mathrm{e}^{\mathrm{i} m \phi_{\mathbf{R}'_{lj}}}}_{\mathrm{singular \ CWs}}\,,
\end{align}
where $\sum_{l,j,m} \equiv \sum_{l = 1}^{N_{x'}} \sum_{j = 1}^{N_{y'}} \sum_{m = -1}^1$, $\mathbf{R}'_{lj} = (x' - x'_l,y' - y'_j, 0)$, $\phi_{\mathbf{R}'_{lj}} =  \arctan\left[\frac{y' - y'_j}{x' - x'_l}\right]$, and the expansion coefficients are $\mathcal{B}_{l,j,m} = \mathrm{e}^{-\mathrm{i} k x'_l} (-\mathrm{i})^m$~\cite{Abramowitz1965Jan}. 

Next, we assemble coefficients $\mathcal{A}_{l,j,m}$ and $\mathcal{B}_{l,j,m}$ into vectors $\boldsymbol{\mathcal{A}}$ and $\boldsymbol{\mathcal{B}}$. The $T$-matrix maps coefficients $\boldsymbol{\mathcal{B}}$ to unknown coefficients $\boldsymbol{\mathcal{A}}$:
$
    \boldsymbol{\mathcal{A}} = \boldsymbol{\mathcal{T}}  \boldsymbol{\mathcal{B}}\, $ where $ \boldsymbol{\mathcal{T}} = \left[\mathbf{I}  - \boldsymbol{\mathcal{C}}\boldsymbol{\mathcal{T}}_{\rm array} \right]^{-1}\boldsymbol{\mathcal{T}}_{\rm array}\,.
$
Here, we have constructed a block-diagonal matrix $\boldsymbol{\mathcal{T}}_{\rm array}$ from the matrix in Eq.~(\ref{eq_Tchain_cw}) repeated $N_{x'} \times N_{y'}$ times and matrix $\boldsymbol{\mathcal{C}}$ from the singular translation coefficients that expand singular CWs in regular CWs at a different position $\mathbf{R}''\neq \mathbf{R}'$~\cite{Abramowitz1965Jan}. Under normal incidence, these coefficients are given by
$
    H_m^{(1)}(k|\mathbf{R}'|) \mathrm{e}^{\mathrm{i} m \phi_{\mathbf{R}'}} = \sum_{m'=-1}^{1} H_{m-m'}^{(1)}(k|\mathbf{R}' - \mathbf{R}''|) \mathrm{e}^{\mathrm{i} (m - m') \phi_{\mathbf{R}'-\mathbf{R}''}} J_{m'}(k|\mathbf{R}''|) \mathrm{e}^{\mathrm{i} m' \phi_{\mathbf{R}''}}\,.
$

\noindent\textbf{Calculation of the effective parameters}

Assuming the impedance matching condition, we calculate the effective speed of sound as $v_i^{\rm eff} = \left( \frac{\Delta \phi_i}{2 \pi f_0 d_{x'}} + v_0^{-1}\right)^{-1}$, where $\Delta \phi_i$ is a phase shift of the total field $p\left(x'_{\rm e},y'_{\rm e}\right)$ [Eq.~\eqref{eq_pressure}] with respect to the incident field $p_{\rm inc}\left(x'_{\rm e},y'_{\rm e}\right)$ [Eq.~\eqref{eq_pressure_inc}] at the evaluation point $\mathbf{r}'_{\rm e} = \left(x'_{\rm e},y'_{\rm e}\right)$. The effective density is then obtained as $\rho_i^{\rm eff} = \rho_0 v_0/v_i^{\rm eff}$. The coordinates were $(-\lambda, 0)$ for the case $x'=x$ and $x'=y$ shown in Fig.~\ref{fig:SI_Fig_2}B. Please note that the incident beam propagates along the $(-x')$-axis, where $x'= i$ and $i=x,y,z$ denotes one of the microcrack axes. 

\noindent\textbf{Computation of corrections to the effective density}

Next, we evaluate corrections to the effective density arising from finite reflection. To this end, we consider a slab of $N_{z'} = 45$ layers of microcracks stacked along the $z'$-axis, where each layer is an infinite, two-dimensional, and periodic lattice of microcracks in the $x'y'$-plane. The \textit{T}-matrix-based approach for computing the response of a two-dimensional lattice of acoustic scatterers in a spherical-wave basis is detailed in Ref.~\cite{Ustimenko2025Oct}. 

We then calculate the scattering matrix (\textit{S}-matrix) of a single lattice layer, $\mathbf{S}_{\rm lattice}$, which contains the transmission and reflection coefficients in the plane-wave basis, as well as the \textit{S}-matrix corresponding to free propagation in the glass medium over distance $s_{z'}$, $\mathbf{S}_{\rm free}$. The \textit{S}-matrix of the multilayered structure is subsequently obtained using the Redheffer star product~\cite{Redheffer1959}
\begin{align}
    \mathbf{S} = \underbrace{\mathbf{S}_{\rm lattice} \star \mathbf{S}_{\rm free} \hdots \star \mathbf{S}_{\rm lattice} \star \mathbf{S}_{\rm free}}_{\left(N_{z'} - 1\right) \, \text{times}} \star \, \mathbf{S}_{\rm lattice}\,,
\end{align}
where the matrix $\mathbf{S}$ contains the transmission ($t$) and reflection ($r$) coefficients of the slab~\cite{Sainidou2005Mar}. 

The plane-wave basis should include a sufficiently large number of diffraction orders to describe near-field coupling between layers. In our simulations, it includes diffraction orders corresponding to reciprocal-lattice vectors $\mathbf{G}$ whose length is $|\mathbf{G}| \leq b_{\rm max}$, where $b_{\rm max}$ was chosen as $b_{\rm max} s_{z'} / (2 \pi) = 3.1$ to ensure convergence of $r$. Further, under $r$,  we assume the reflection coefficient of the zeroth diffraction order. 

Within the \textit{S}-matrix formalism, the incident field propagates along the $z'$-axis, which coincides with one of the microcrack axes $i$. From the slab reflection coefficient $r_i$, the effective reflection coefficient of the glass--metamaterial interface, $r_i^{\rm eff}$, can be obtained by solving Eq.~\eqref{eq:r_eff} with $v^{\rm eff}_x = 5084$ m s$^{-1}$, $v^{\rm eff}_y = 5615$ m s$^{-1}$, and $v^{\rm eff}_z = 5687$ m s$^{-1}$. Substituting $r_i^{\rm eff}$ into Eq.~\eqref{eq:delta_rho} results in the correction to the effective density, $\delta \rho_i^{\rm eff}$.

The theoretically computed reflection coefficients and density corrections are also included in Table~\ref{tab:table_density_num}. Their agreement with the experimentally obtained values further supports the assumption of impedance matching.

\noindent\textbf{Influence of shear waves}

In the \textit{T}-matrix-based model of the metamaterial response, shear waves were neglected. To justify this approximation, we analyzed the impact of shear waves on the effective parameters. To this end, we performed a simulation in COMSOL Multiphysics\texttrademark, v.6.3, based on the finite-element method (FEM). In the FEM models, we consider the metamaterial as a $45\times45$ array of circular cylinders that are infinitely extended along the laser direction ($z'=z$) and whose radii are equal to $a_x$. For simplicity, we assume that the cylinders are isotropic. Under these assumptions, the problem becomes effectively two-dimensional (2D). The material parameters of the cylinders (effective density and bulk modulus) were retrieved from homogenization of a microcrack chain along the $z'$-axis as follows. First, we calculated the quasistatic monopole and dipole polarizabilities of a cylinder using the elements $\mathcal{T}_{m,m'}$ of $\boldsymbol{\mathcal{T}}_{\rm chain}$ in Eq.~\eqref{eq_Tchain_cw}~\cite{Torrent2006Dec}:
\begin{align}
\alpha^{\rm cyl}_{\rm M}  = \frac{4}{\mathrm{i}k^2} \left(\mathcal{T}_{0,0}^{-1} + 1\right)^{-1} \,, \quad \text{and} \quad
\boldsymbol{\alpha}^{\rm cyl}_{\rm D} = \frac{8}{\mathrm{i}k^2}
\left[
\mathbf{F}_{\rm 2D} \begin{pmatrix}
\mathcal{T}_{\mathrm{-1,-1}} & \mathcal{T}_{\mathrm{-1,1}} \\
\mathcal{T}_{\mathrm{1,-1}} & \mathcal{T}_{\mathrm{1,1}}
\end{pmatrix}^{-1}\mathbf{F}^{-1}_{\rm 2D} + \mathbf{I}_2
\right]^{-1}\,,
\end{align}
respectively. Matrices $\mathbf{F}_{\rm 2D}$ and $\mathbf{F}^{-1}_{\rm 2D}$ convert the Cartesian components of the dipole moment into the cylindrical-wave basis, and vice versa, respectively,
\begin{align}
\mathbf{F}_{\rm 2D} = \frac{1}{\sqrt{2}}\begin{pmatrix}
1& -1\\
-\mathrm{i}& -\mathrm{i}
\end{pmatrix}\,,
\quad \text{and} \quad
\mathbf{F}^{-1}_{\rm 2D} = \mathbf{F}^{\dagger}_{\rm 2D} =  \frac{1}{\sqrt{2}}\begin{pmatrix}
1& -\mathrm{i}\\
-1& -\mathrm{i}
\end{pmatrix}\,.
\end{align}
Next, we calculated the normalized material parameters from the static polarizabilities as
\begin{align}
    \widetilde{\beta}^{\rm cyl} = 1 + \alpha^{\rm cyl}_{\rm M}/A_{\rm cyl}\,, \quad \text{and} \quad \widetilde{\rho}^{\rm cyl} = \frac{2 + \alpha^{\rm cyl}_{{\rm D}, xx}/A_{\rm cyl}}{2 - \alpha^{\rm cyl}_{{\rm D}, xx}/A_{\rm cyl}}\,,
\end{align}
where $A_{\rm cyl} = \pi a_x^2$ is the cross section area of the cylinder. For the considered system, the following parameters were obtained: $\widetilde{\beta}^{\rm cyl} = 1.46$ and $\widetilde{\rho}^{\rm cyl} = 4.31$.

The Poisson's ratio of the glass medium has been set to 0.206, which corresponds to the value for BK7 glass~\cite{su_refractive_2008}. The same Poisson’s ratio was also assumed for the microcracks in a numerical model with shear waves created in the \textit{Solid Mechanics} interface, while the \textit{Pressure Acoustics} interface was used for a numerical model without shear waves. Both models accounted for attenuation in the glass medium at 2.25 MHz (Table~\ref{tab:tab2}). As the incident field, a plane pressure wave was considered. From a phase shift of the transmitted wave, we estimated the effective speed of sound of the metamaterial using the same approach as in the \textit{T}-matrix-based model. All predicted values of the effective speed of sound lie within the confidence interval of the measured value $v^{\rm eff}_x/v_0 = (0.83\pm0.07)$ and differ only by $\approx$1\% (see Table~\ref{tab:table_shear}). Moreover, these results agree with the values of the effective speed of sound computed by the \textit{T}-matrix-based model, further confirming the validity of the no-shear-wave approximation.

\subsubsection*{Optimization of the laser engraving step}

\begin{figure}[h!]
	\centering
	\includegraphics[width=1\textwidth]{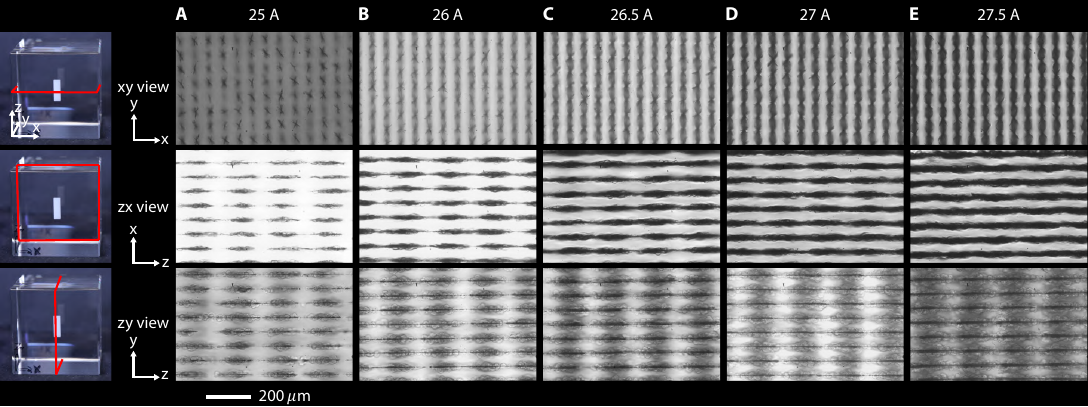}
	\caption{\textbf{Microscopy images of engraved microcracks for different laser currents.}
		Five samples were engraved with different laser diode currents of the pulsed laser: (\textbf{A}) 25 A, (\textbf{B}) 26 A, (\textbf{C}) 26.5 A, (\textbf{D}) 27 A, (\textbf{E}) 27.5 A. Three orthogonal views outlined in red ($xy$-top panel, $zx$-middle panel, and $zy$-bottom panel) were imaged with a bright-field microscope. The $z$-axis is the direction of the engraving laser. The size of the microcracks increases with the laser diode current.}
	\label{fig:SI_Fig_1}
\end{figure}

\begin{figure}[h!]
	\centering
	\includegraphics[width=0.8\textwidth]{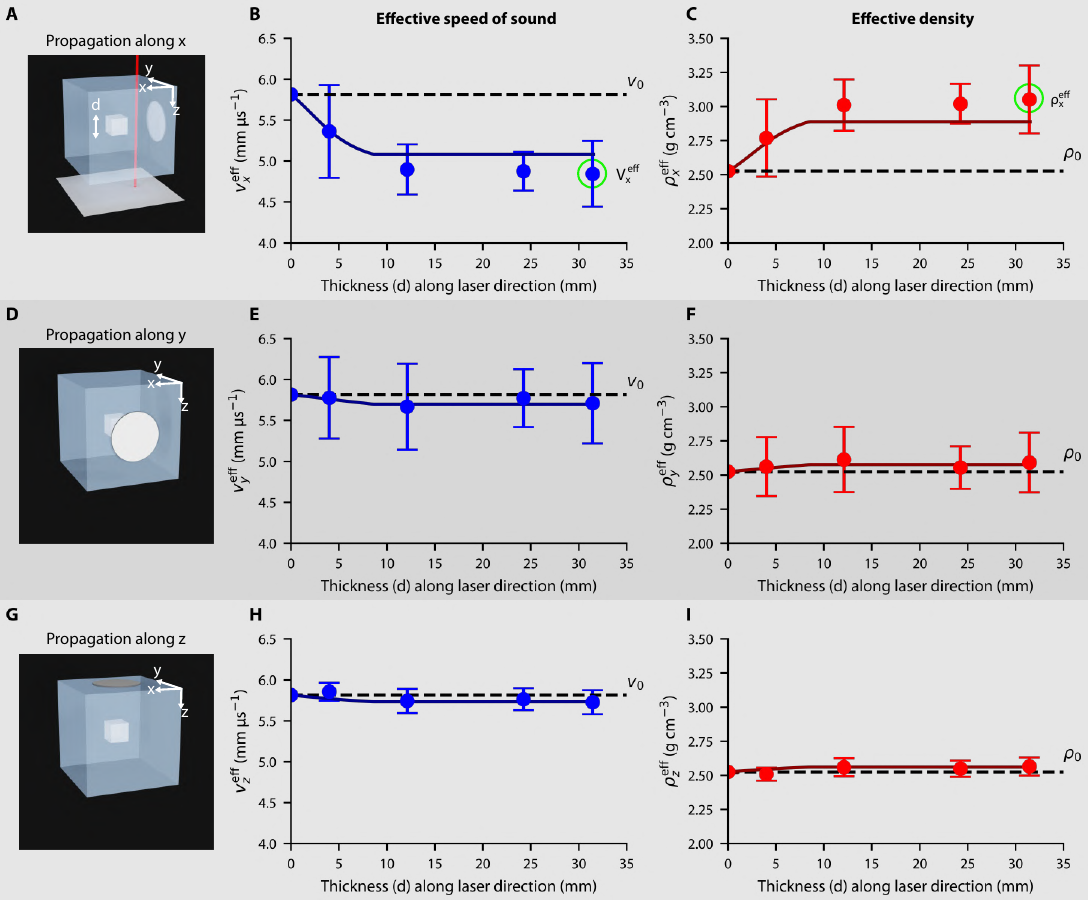}
	\caption{\textbf{Effective longitudinal speed of sound: \textit{T}-matrix simulation and measurement.}
    With refracto-vibrometry measurements, a 3D sound field is reduced to a 2D projection. Therefore, the size of the engraved voxel along the vibrometry laser direction relative to the block size affects the measurement. The effective speed of sound and density were measured for different propagation directions (along $x$, $y$, or $z$). (\textbf{A}) Schematics of the setup for propagation along the $x$-axis. (\textbf{B}) Effective speed of sound $v^{\rm eff}_x$ for different sizes of the engraved voxel $d$ depicted in panel \textbf{A}. The \textit{T}-matrix simulation data are depicted by a line, experimental refracto-vibrometry measurements by dots. The plot reaches a plateau value, which is taken as the final value of $v^{\rm eff}_x$ (green circle). For comparison, the speed of sound in the bulk medium $v_0$ is plotted with a horizontal dotted black line. (\textbf{C}) Effective density $\rho^{\rm eff}_x$ for different sizes of the engraved voxel. The plot reaches a plateau value, which is taken as the final value of $\rho^{\rm eff}_x$ (green circle). For comparison, the density in the bulk medium $\rho_0$ is plotted with a horizontal dotted black line. 
    (\textbf{D})-(\textbf{F}), same as (\textbf{A})-(\textbf{C}) but for the ultrasound propagation along the $y$-axis.
    (\textbf{G})-(\textbf{I}), same as (\textbf{A})-(\textbf{C}) but for the ultrasound propagation along the $z$-axis}
	\label{fig:SI_Fig_2}
\end{figure}

\begin{figure}[h!]
	\centering
	\includegraphics[width=1\textwidth]{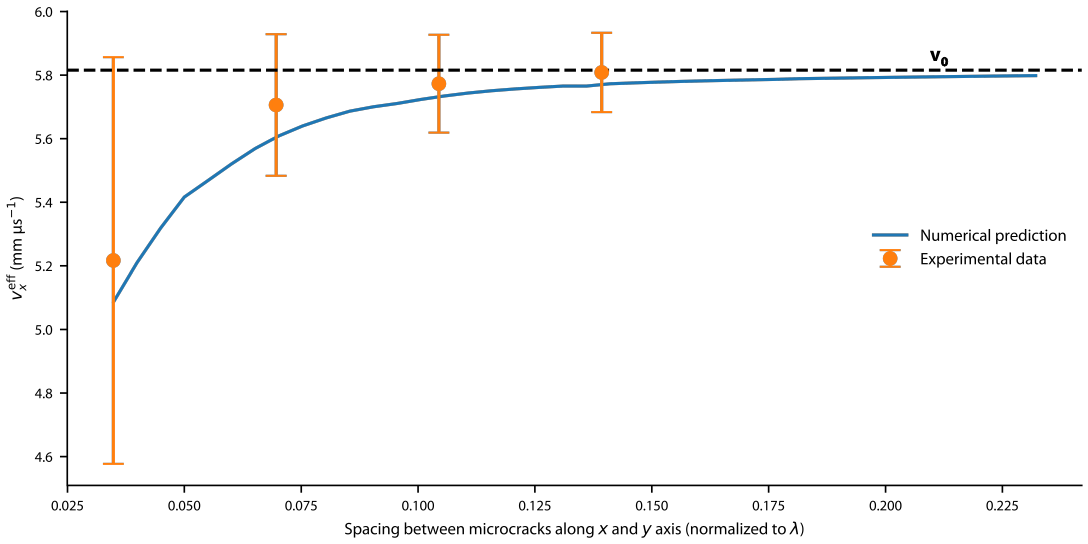}
      	\caption{\textbf{Effective longitudinal speed of sound for different spacings.}
		The effective longitudinal speed of sound along the $x$-axis was measured for engraved samples with different spacings between microcracks along the $x$ and $y$-axes. Experimental data points (orange markers) confirm the prediction of the \textit{T}-matrix method (blue solid line). The longitudinal speed of sound in the bulk medium $v_{0}$ is depicted by the horizontal black dashed line.}
	\label{fig:SI_Fig_3}
\end{figure}

\begin{figure}[h!]
	\centering
	\includegraphics[width=1\textwidth]{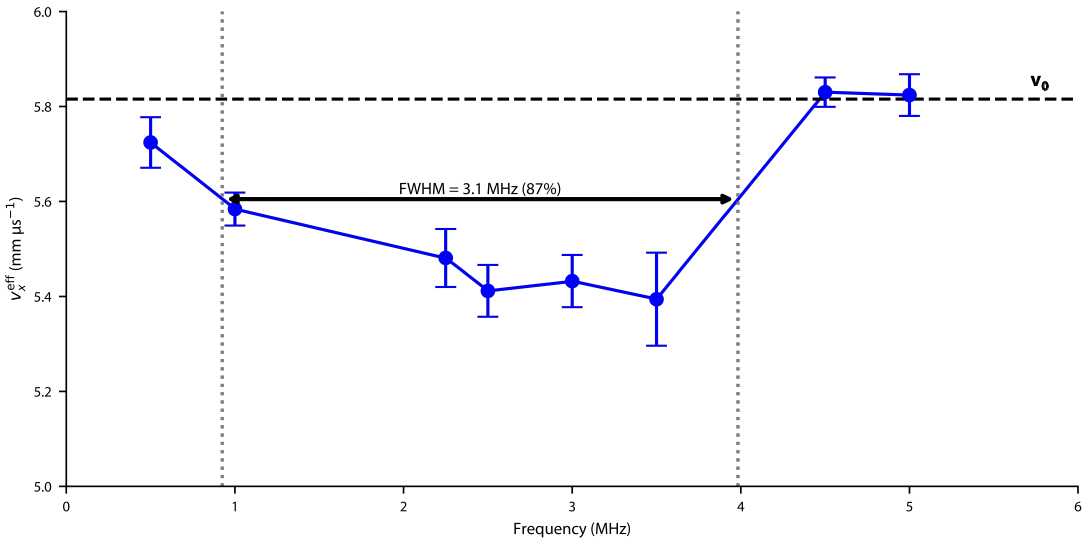}
      	\caption{\textbf{Frequency-selective response of the metamaterial.}
		The effective longitudinal speed of sound $v^{\rm eff}_{x}$ along the $x$-axis was measured for different frequencies using four transducers, whose center frequencies are 0.5, 1, 2.25, and 5 MHz, respectively.  The longitudinal speed of sound in the bulk medium $v_{0}$ is depicted by the horizontal black dashed line. The measurement of the full width at half maximum (FWHM) gives a fractional bandwidth of 87\%, shown by the vertical dotted gray lines.}
	\label{fig:SI_Fig_4}
\end{figure}

\begin{figure}[h!]
	\centering
	\includegraphics[width=1\textwidth]{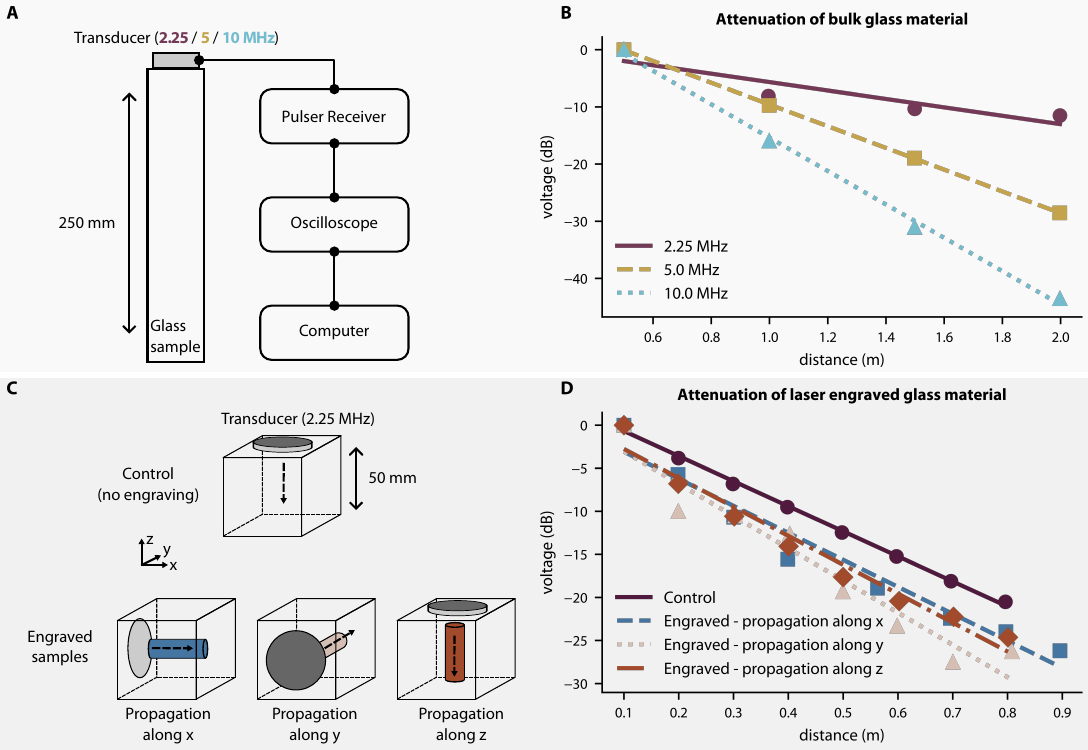}
	\caption{\textbf{Attenuation measurement of the bulk and engraved glass medium.}
        The attenuation inside the bulk medium (top panel) and inside the engraved medium (bottom panel) was measured. (\textbf{A}) Schematics of the experimental setup with a bigger glass block of size $70 \times 70 \times 250$ mm and a pulse-echo measurement. (\textbf{B}) The maximum absolute value of the signal in dB is plotted after different reflections, i.e., different propagation distances. The amplitude decreases linearly with the distance at 2.25, 5, and 10 MHz. (\textbf{C}) Three glass cubes of size 50 mm were then engraved with a cylindrical pattern oriented along the $x$, $y$, and $z$-directions (with the $z$-axis being the direction of the engraving laser). The attenuation was measured for each engraved block and for the control block of the same size. (\textbf{D}) At 2.25 MHz, a 10\% higher attenuation is observed for the engraved blocks.}
	\label{fig:SI_Fig_5}
\end{figure}

\begin{figure}[h!]
	\centering
	\includegraphics[width=0.6\textwidth]{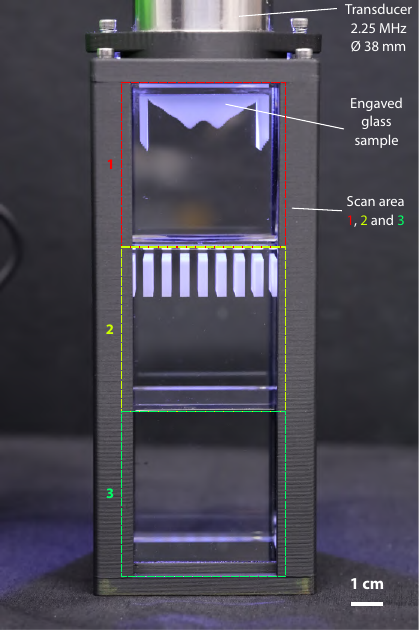}
	\caption{\textbf{Photograph of the setup for the stacking of blocks.}
        A 3D-printed scaffold aligns the stacking of glass blocks. Two screws apply a light compressive pressure between the blocks to reduce the acoustic reflection at the interfaces. Each scan area is outlined by a colored box. The scale bar is 1 cm.}
	\label{fig:SI_Fig_6}
\end{figure}

\begin{figure}[h!]
	\centering
	\includegraphics[width=0.8\textwidth]{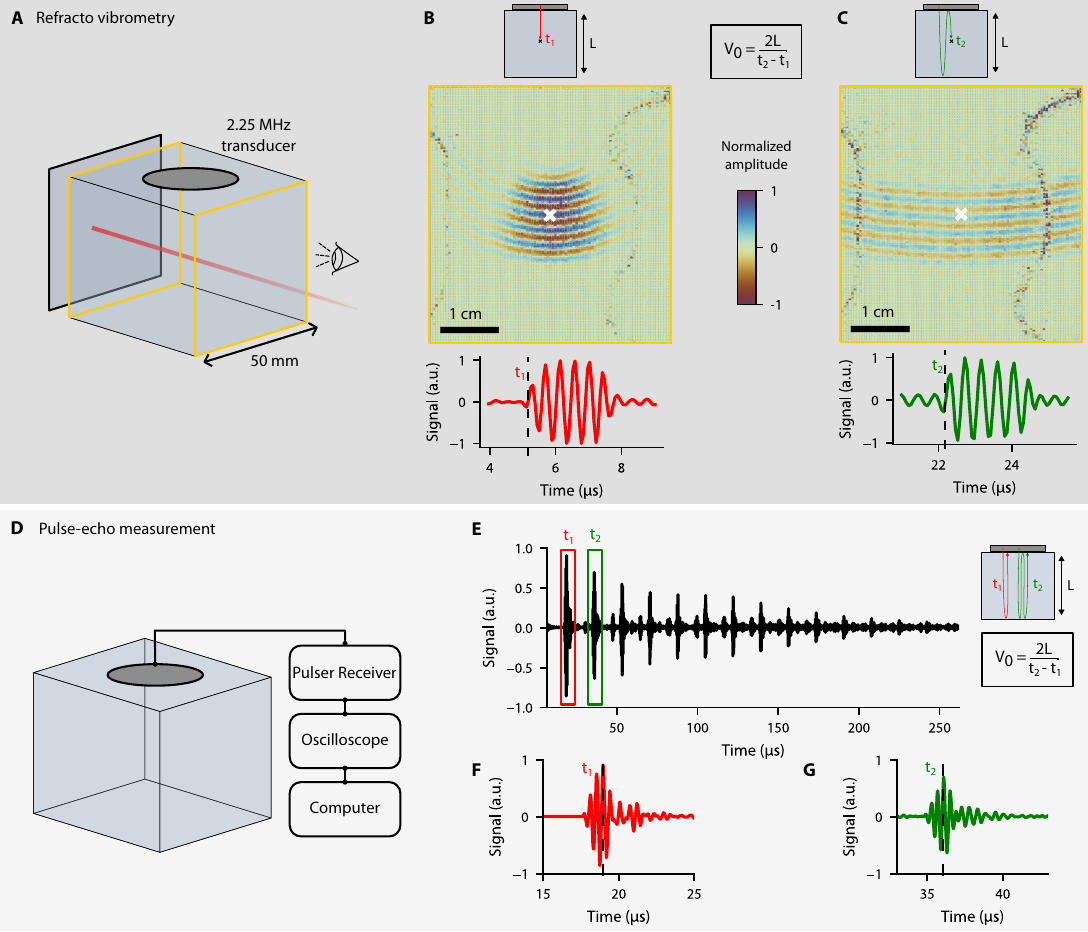}
	\caption{\textbf{Measurement of the speed of sound in the bulk medium.}
        The following two techniques were used: refracto-vibrometry (top panel) and pulse-echo measurement (bottom panel).
		(\textbf{A}) Schematics of the refracto-vibrometry setup. A 3D sound field is reduced to a 2D projection. The eye and the highlighted faces of the cube depict the measurement direction. (\textbf{B}) Measurement of a control block (no engraving). A time profile at a scanning point (white cross) is plotted in red. Time-of-flight $t_{1}$ is indicated. (\textbf{C}) Signal after reflections from the bottom and top faces of the cube. A time profile at the same scanning point is plotted in green. Time-of-flight $t_{2}$ is indicated. The equation used for the speed of sound calculation is outlined in black. 
        (\textbf{D}) Schematics of the pulse-echo setup. (\textbf{E}) Plot of the measured signal over multiple reflections. (\textbf{F}) Subset of the signal (red box) for the first reflection. (\textbf{G}) Subset of the signal (green box) for the second reflection. Times-of-flight $t_{1}$ and $t_{2}$ for both signals are indicated. The equation used for the speed of sound calculation is outlined in black. All scale bars indicate 1 cm.}
	\label{fig:SI_Fig_7}
\end{figure}

\begin{figure}[h!]
	\centering
	\includegraphics[width=1\textwidth]{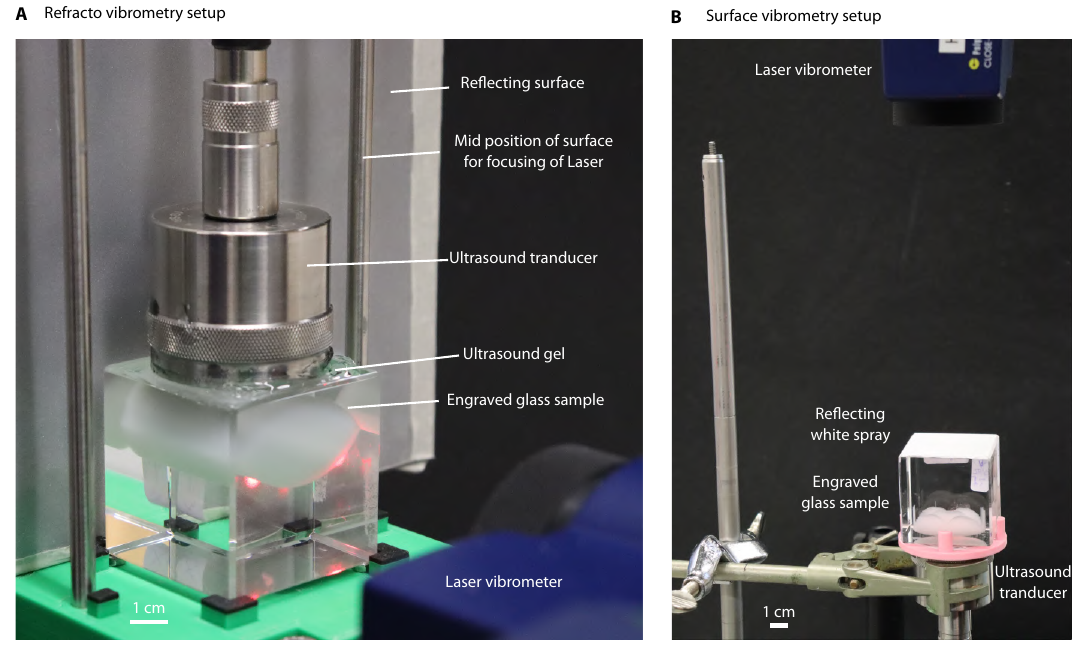}
	\caption{\textbf{Photographs of the vibrometry setups.}
        Two vibrometry setups were used in this study. (\textbf{A}) A refracto-vibrometry setup scans a transparent block of glass while a 2.25 MHz burst is transmitted from the top face. (\textbf{B}) A standard surface vibrometry setup scans the vibration at the surface of a face opposite the transducer. All scale bars indicate 1 cm.}
	\label{fig:SI_Fig_8}
\end{figure}

\begin{figure}[h!]
	\centering
	\includegraphics[width=1\textwidth]{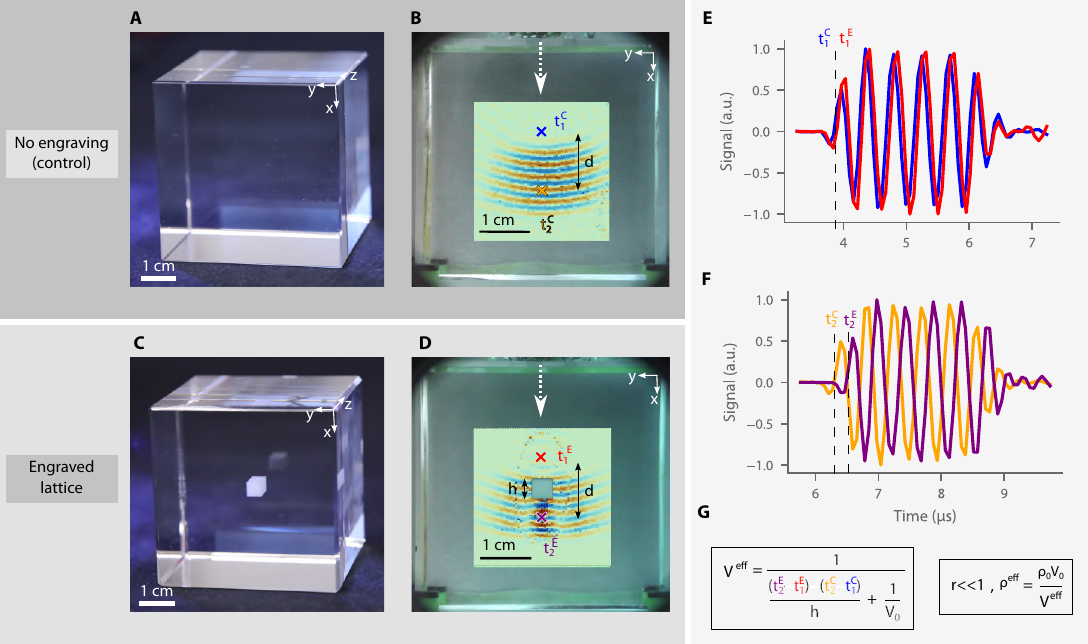}
	\caption{\textbf{Measurement of the effective longitudinal speed of sound.}
    To measure the effective speed of sound, two samples were used: a control block (top panel) and a block with an engraved voxel at its center (bottom panel).
    (\textbf{A}) Photograph of the control block. White arrows indicate the $x$, $y$, and $z$-axes. (\textbf{B}) The refracto-vibrometry signal. A white dotted arrow depicts the propagation direction. (\textbf{C}) Photograph of the engraved block. (\textbf{D}) The same as in panel \textbf{B} but for the engraved block. (\textbf{E}) Overlaid time signal for two scanning points (blue and red crosses). $t_{1}^{\rm C}$ and $t_{1}^{\rm E}$ are the time-of-flight for the control and engraved blocks, respectively. (\textbf{F}) Overlaid time signals for two scanning points (yellow and purple crosses). $t_{2}^{\rm C}$ and $t_{2}^{\rm E}$ are the times-of-flight for the control and engraved blocks, respectively. In this case, a delay is observed when the sound propagates through the engraved voxel. (\textbf{G}) The equations used to calculate the effective speed of sound $v^{\rm eff}$ and the effective density $\rho^{\rm eff}$ are outlined in black. They are functions of the times-of-flight, height of the engraved voxel $h$, and also $\rho_0$ and $v_0$, which are the density and the speed of sound of the bulk medium, respectively. For the density calculation, the pressure reflection coefficient was considered negligibly small $r \approx 0$. All scale bars indicate 1 cm.}
	\label{fig:SI_Fig_9}
\end{figure}

\begin{figure}[h!]
	\centering
	\includegraphics[width=1\textwidth]{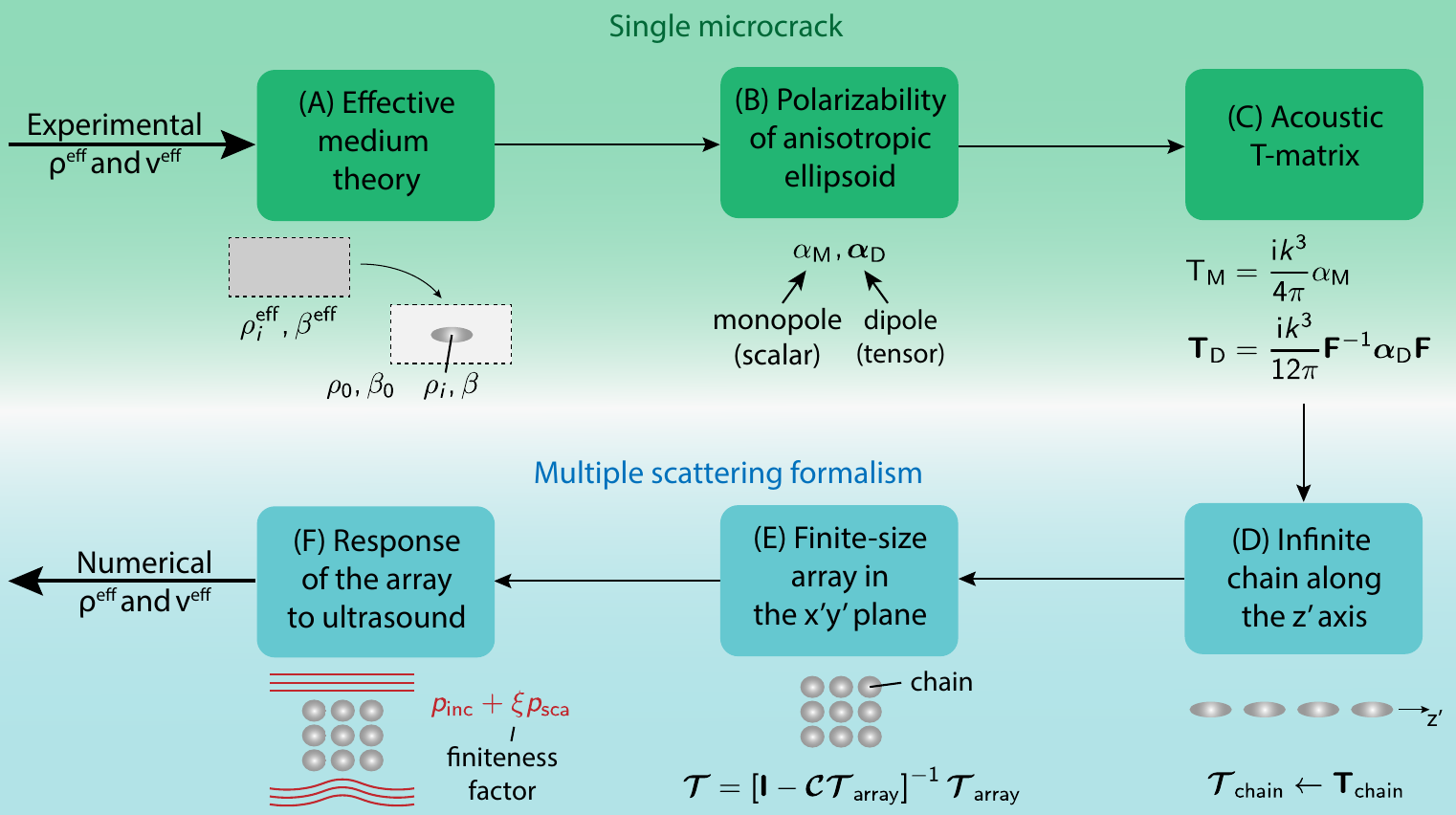}
	\caption{\textbf{Schematics of the \textit{T}-matrix-based numerical approach}. (\textbf{A}) We estimate the material parameters of a microcrack from the measured effective parameters utilizing an effective medium theory for noninteracting, diluted ellipsoids. (\textbf{B}) Using the material parameters, we calculate the monopole and the dipole dynamic polarizabilities of a microcrack in the quasistatic approximation, considering it as an anisotropic ellipsoid, and (\textbf{C}) convert them into the \textit{T}-matrix of an isolated microcrack in the spherical-wave basis. (\textbf{D}) We calculate the \textit{T}-matrix of a periodic chain infinitely extended along the refracto-vibrometry measurement axis ($z^{\prime}$-axis) from the single-microcrack \textit{T}-matrix in the spherical-wave basis, and subsequently transform it into the cylindrical-wave basis. (\textbf{E}) In the $z' = 0$ plane, we consider a microcrack pattern as a finite 2D array of chains and compute the \textit{T}-matrix $\boldsymbol{\mathcal{T}}$ that accounts for the collective interactions between the chains. (\textbf{F}) At the final step, we use this \textit{T}-matrix to map the expansion coefficients of the incident field to those of the scattered field for each chain. The total scattered field is the superposition of the fields scattered by all chains. It is then multiplied by the prefactor $\xi$ that accounts for the finite size of the engraved area along the $z'$-axis and for the finite radius of the beam waist the top face of the engraved area.}
	\label{fig:SI_Fig_10}
\end{figure}

\begin{figure}[h!]
	\centering
	\includegraphics[width=1\textwidth]{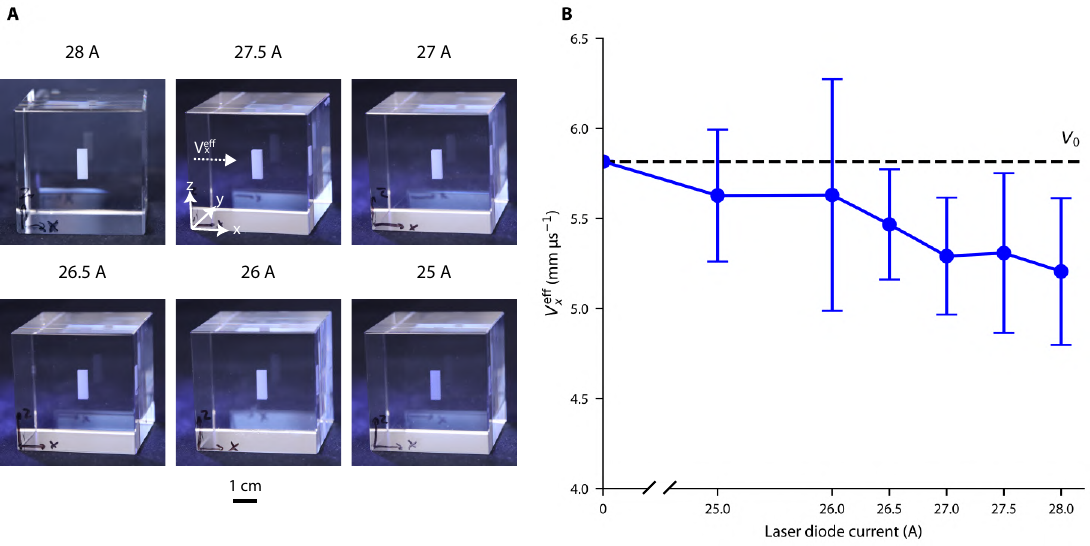}
	\caption{\textbf{Measurement of effective speed of sound for different laser diode currents.}
		Six samples were engraved with different laser diode currents of the pulsed laser. 
        (\textbf{A}) Pictures of the different samples. White arrows indicate the $x$, $y$, and $z$-axes. A white dotted arrow depicts the propagation direction. (\textbf{B}) The effective speed of sound $v^{\rm eff}_{x}$ for the $x$-axis was measured for the different laser diode currents. For comparison, the speed of sound in the bulk medium $v_0$ is depicted by the horizontal dotted black line. The effective speed of sound decreases with the laser diode current.}
	\label{fig:SI_Fig_11}
\end{figure}

Several parameters were optimized to prevent the glass from shattering and the failure of the engraving process. First, the spacing between points was set to the minimum spacing [90, 90, 270] \textmu m along the $x$, $y$, and $z$-axes, respectively. For smaller spacings, the microcracks would merge and lead to major cracks. The spacing is bigger along the $z$-axis because the microcracks are longer along this direction. Then, a safety margin of 4 mm was used between the side faces of the glass samples and the microcracks. Eventually, the laser diode current was set to $27.5$ A. At higher values, the engraving process was more likely to fail because of larger cracks emerging through the sample. At lower values, the engraved microcracks were too small (Fig.~\ref{fig:SI_Fig_1}) and the acoustic parameters of the effective medium were almost the same as those of the bulk medium (Fig.~\ref{fig:SI_Fig_11}). \\ Potentially, any 3D shape could be engraved as long as the engraving parameters are close to the ones described above.





\begin{table}[h!]
	\centering
	\caption{\textbf{Thermal properties of VeroClear and BK7 glass.}
		Comparison of thermal properties for VeroClear~\cite{bakaric_measurement_2021, ulu_development_2020, vaithilingam_optimisation_2018} and BK7 glass~\cite{su_refractive_2008}.}
	\label{tab:table_s1}
    \begin{tabular}{c|c|c|}
    \cline{2-3}
     & VeroClear & BK7 glass \\ \hline
    \multicolumn{1}{|c|}{Glass transition temperature (°C)} & 54 & 557 \\ \hline
    \multicolumn{1}{|c|}{Density (kg\,m$^{-3}$)} & 1180 & 2510 \\ \hline
    \multicolumn{1}{|c|}{Specific heat capacity (J\,kg$^{-1}$\,K$^{-1}$)} & 1349 & 858 \\ \hline
    \multicolumn{1}{|c|}{Volumetric heat capacity (MJ\,K$^{-1}$\,m$^{-3}$)} & 1.59 & 2.15 \\ \hline
    \multicolumn{1}{|c|}{Thermal conductivity (W\,m$^{-1}$\,K$^{-1}$)} & 0.2261 & 1.1 \\ \hline
    \multicolumn{1}{|c|}{Thermal expansion coefficient (µm$^{-1}$\,m$^{-1}$\,°C$^{-1}$)} & 200 & 5.6 \\ \hline
\end{tabular}
\end{table}

\begin{table}[h!]
	\centering
	\caption{\textbf{Parameters for the refracto-vibrometry measurements.}
		The setup consists of a Polytec PSV 500 laser vibrometer, a signal generator, an amplifier, and an ultrasound transducer.}
	\label{tab:table_s2}
    \begin{tabular}{|c|c|c|}
    \hline
    Vibrometer & Scanning head & PSV-I-500 \\ \cline{2-3} 
     & Laser wavelength & 632.8 nm \\ \cline{2-3} 
     & Acquisition mode & Time \\ \cline{2-3} 
     & Sample length & 500 \\ \cline{2-3} 
     & Sample frequency (MHz) & 15.625 \\ \cline{2-3} 
     & Sample time (us) & 32 \\ \cline{2-3} 
     & Resolution (ns) & 64 \\ \cline{2-3} 
     & Optics & \begin{tabular}[c]{@{}c@{}}Close-up unit \\ + micro scan lens (PSV-A-CL-200-HeNe)\end{tabular} \\ \cline{2-3} 
     & Field of view of the scanning area (mm) & ~40 x 40 \\ \cline{2-3} 
     & Spacing between scanning points & ~ 0.5 mm = $\lambda/5$ \\ \hline
    Signal generator & Signal & 5-cycle burst centered at 2.25 MHz \\ \cline{2-3} 
     & Voltage (V) & ~0.2 \\ \cline{2-3} 
     & Pulse Repetition Frequency (kHz) & 1 \\ \hline
    Amplifier & Frequency range & 10 kHz to 12 MHz \\ \cline{2-3} 
     & Power gain (dB) & 50 \\ \cline{2-3} 
     & Input and output impedance (ohms) & 50 \\ \hline
    Transducer & Center frequency (MHz) & 2.25 \\ \cline{2-3} 
     & Diameter (mm) & 38.1 mm \\ \hline
    \end{tabular}
\end{table}

\begin{table}[h!]
	\centering
	\caption{\textbf{Comparison of existing 3D metamaterials for MHz ultrasonic waves}.}
	\label{tab:table_literature_3d}
    \resizebox{\linewidth}{!}{
    \begin{tabular}{|c|c|c|c|c|c|c|c|}
    \hline
    Reference&  \begin{tabular}[c]{@{}c@{}}Frequency \\ (MHz)  \end{tabular}  &Function &\begin{tabular}[c]{@{}c@{}}Aperture \\ (cm)   \end{tabular}  &\begin{tabular}[c]{@{}c@{}}Fabrication \\ method \end{tabular} &\begin{tabular}[c]{@{}c@{}}Bulk \\ material  \end{tabular} &\begin{tabular}[c]{@{}c@{}}Attenuation \\ of bulk material \\ (dB cm$^{-1}$ MHz$^{-1}$)   \end{tabular} \\\hline
    Our work & $\approx 2$ & \begin{tabular}[c]{@{}c@{}}Phase \\ modulation \\ over \\ 
    $2\pi$ range   \end{tabular} & 5 & \begin{tabular}[c]{@{}c@{}}Subsurface \\ laser \\ engraving   \end{tabular} & BK7 glass & 0.03\\ \hline
    Ref.~\cite{brown_binary_2023}&  $\approx 2$ & \begin{tabular}[c]{@{}c@{}}Phase \\ modulation \\ over \\ $2\pi$ range   \end{tabular} & 3 & \begin{tabular}[c]{@{}c@{}}3D \\ printing   \end{tabular} & \begin{tabular}[c]{@{}c@{}}VeroClear \\ (PMMA-like)    \end{tabular} & $\approx 3.7$\\ \hline
    Ref.~\cite{IglesiasMartinez2021Feb}& $1-8$ & \begin{tabular}[c]{@{}c@{}}Bandgap \\ control \end{tabular} & 0.1 & \begin{tabular}[c]{@{}c@{}}Microscale \\ 3D fabrication \\ with two-photon \\ lithography    \end{tabular}& \begin{tabular}[c]{@{}c@{}}Ip-S \\ photoresist    \end{tabular} & $\approx 6$~\cite{Kim2023May} \\ \hline
    \end{tabular}
    }
\end{table}

\begin{table}[h!]
	\centering
	\caption{\textbf{Comparison of existing 2D metamaterials for MHz ultrasonic waves}.}
	\label{tab:table_literature_2d}
    \resizebox{\linewidth}{!}{
    \begin{tabular}{|c|c|c|c|c|c|c|c|}
    \hline
    Reference& \begin{tabular}[c]{@{}c@{}}Frequency \\ (MHz)  \end{tabular}  &Function &\begin{tabular}[c]{@{}c@{}}Aperture \\ (cm)   \end{tabular}   &\begin{tabular}[c]{@{}c@{}}Fabrication \\ method \end{tabular} &\begin{tabular}[c]{@{}c@{}}Bulk \\ material  \end{tabular} &\begin{tabular}[c]{@{}c@{}}Attenuation \\ of bulk material \\ (dB cm$^{-1}$ MHz$^{-1}$)   \end{tabular} \\\hline
    Ref.~\cite{xu_sound-speed_2023}& $\approx 6.9$ & \begin{tabular}[c]{@{}c@{}}Phase \\ modulation \\ over \\ $2\pi$ range   \end{tabular}& 1.5 & \begin{tabular}[c]{@{}c@{}}Photolithography \\ \& etching   \end{tabular} & \begin{tabular}[c]{@{}c@{}}Silicon \\ wafer   \end{tabular}& $\approx 0.08$~\cite{Sylwestrowicz1966Feb} \\ \hline
    Ref.~\cite{Dorn2026Feb}& $0.7-1$ & \begin{tabular}[c]{@{}c@{}}Waveguide \\ control \\ through \\ dispersion \\ engineering   \end{tabular}& 3 & \begin{tabular}[c]{@{}c@{}} Photolithography \\ \& etching \end{tabular}&
\begin{tabular}[c]{@{}c@{}}Silicon \\ wafer   \end{tabular}& $\approx 0.08$~\cite{Sylwestrowicz1966Feb} \\ \hline
Ref.~\cite{Olsson2008} &  $60-75$ & \begin{tabular}[c]{@{}c@{}}Bandgap \\ \& waveguide \\ control   \end{tabular}& 0.1 & \begin{tabular}[c]{@{}c@{}} Photolithography \\ \& etching \end{tabular} & \begin{tabular}[c]{@{}c@{}}SiO$_2$  \\host matrix   \end{tabular}& $\approx 0.08$~\cite{Sylwestrowicz1966Feb} \\ \hline
Ref.~\cite{jin_flat_2019} &  $\approx0.2$ & \begin{tabular}[c]{@{}c@{}}Phase \\ modulation \\ over \\ $2\pi$ range   \end{tabular} & 5 & \begin{tabular}[c]{@{}c@{}}Manual \\ alignment \\ of porous \\ silicone \\ rubber \\ stripes \end{tabular}& PDMS & $\approx1.7$~\cite{Xu2020May} \\ \hline
Ref.~\cite{melde_holograms_2016} &  $\approx 2$ & \begin{tabular}[c]{@{}c@{}}Phase \\ modulation \\ over \\ $2\pi$ range   \end{tabular} & 5 & \begin{tabular}[c]{@{}c@{}}3D printing \end{tabular}& \begin{tabular}[c]{@{}c@{}}VeroClear \\ (PMMA-like)    \end{tabular} & $3.7$ \\ \hline
Ref.~\cite{attali_deep_2025} & $0.5$ & \begin{tabular}[c]{@{}c@{}}Phase \\ modulation \\ over \\ $2\pi$ range   \end{tabular} & 10 & \begin{tabular}[c]{@{}c@{}}3D printing \\ \& molding \end{tabular}& Silicone & $4$ \\ \hline
Ref.~\cite{Melnikov2022Jul}&  $\approx 2$ & \begin{tabular}[c]{@{}c@{}}Anomalous \\ refraction  \end{tabular} & 0.3 & \begin{tabular}[c]{@{}c@{}}Two-photon \\ polymerization \end{tabular}& \begin{tabular}[c]{@{}c@{}}Ip-S \\ photoresist \end{tabular} & $\approx 6$~\cite{Kim2023May} \\ \hline
    \end{tabular}
    }
\end{table}

\begin{table}[h!]
\centering
\caption{\textbf{Effect of reflection on the measurement of the effective density}.  $\delta \rho^{\rm eff}_i$ is the correction obtained when taking into account the effective reflection coefficient of the glass-metamaterial interface $r_i^{\rm eff}$ from the reflection coefficient of a slab (metamaterial) $r_i$. The table compares the results from the \textit{T}-matrix method (theory) with experimental values obtained through refracto-vibrometry. $\rho_0$ is the density of the bulk medium.}
\label{tab:table_density_num}
\resizebox{\columnwidth}{!}{
\begin{tabular}{c|c|c|c|c|c|}
\cline{2-6}
                                 & Propagation axis $i$ & $\rho^{\rm eff}_i/\rho_0$& $\delta \rho^{\rm eff}_i/\rho_0$   & $r_i$                          & $r_i^{\rm eff}$                \\ \hline  
\multicolumn{1}{|c|}{}           & $x$                  & $1.21\pm0.10$              & $0.10\pm0.10$                      & $0.07+\mathrm{i}0.02\pm0.06$   & $0.04-\mathrm{i}0.004\pm0.21$  \\ \cline{2-6} 
\multicolumn{1}{|c|}{Experiment} & $y$                  & $1.03\pm0.09$              & $0.01\pm0.02$                      & $0.01-\mathrm{i}0.002\pm0.01$  & $0.01+\mathrm{i}0.001\pm0.11$  \\ \cline{2-6} 
\multicolumn{1}{|c|}{}           & $z$                  & $1.02\pm0.03$              & $-0.01\pm0.03$                     & $-0.01-\mathrm{i}0.001\pm0.04$ & $-0.01-\mathrm{i}0.001\pm0.09$ \\ \hline
\multicolumn{1}{|c|}{}           & $x$                  & $1.14$                     & $0.05$                             & $0.051-\mathrm{i}0.006$         & $0.025-\mathrm{i}0.007$         \\ \cline{2-6} 
\multicolumn{1}{|c|}{Simulation} & $y$                  & $1.04$                     & $-0.04$                            & $-0.012+\mathrm{i}0.020$         & $-0.019+\mathrm{i}0.002$        \\ \cline{2-6} 
\multicolumn{1}{|c|}{}           & $z$                  & $1.02$                     & $-0.03$                            & $-0.028-\mathrm{i}0.001$        & $-0.014+\mathrm{i}0.002$        \\ \hline
\end{tabular}
}
\end{table}

\begin{table}[h!]
	\centering
	\caption{\textbf{Impact of shear waves on the measurement of the effective speed of sound}. The effective speed of sound was measured experimentally and predicted by FEM models, which are based on the following interfaces of COMSOL Multiphysics\texttrademark, and by the \textit{T}-matrix model. $v_0$ is the speed of sound of the bulk medium.}
	\label{tab:table_shear}
    \begin{tabular}{c|c|c|c|c|}
    \cline{2-5}
     & Experiment & Solid Mechanics & Pressure Acoustics & \textit{T}-matrix \\ \hline
    \multicolumn{1}{|c|}{Effective speed of sound} & $0.83 v_0 \pm 0.07 v_0$& 0.86$v_0$ & 0.87$v_0$ & 0.87$v_0$ \\ \hline
\end{tabular}
\end{table}


\clearpage 





\end{document}